\newif\ifARXIV
\ARXIVtrue

\ifARXIV
\documentclass[journal, a4paper]{IEEEtran}

\usepackage{graphicx}     
\usepackage{listings}

\usepackage{amsfonts}  

\usepackage{hyperref}

\usepackage{hyphenat}
\begin{document}
\bstctlcite{IEEEexample:BSTcontrol}

\title{Survey of Crosschain Communications Protocols}
\author{
     \IEEEauthorblockN{
    	Peter Robinson\\
         } 
    \IEEEauthorblockA{ConsenSys Software R\&D\\}
    \IEEEauthorblockA{School of Information Technology and Electrical Engineering, University of Queensland, Australia\\}
    \IEEEauthorblockA{
    	peter.robinson@consensys.net\\
	}
}

\maketitle

\else
\documentclass{elsarticle}

\usepackage[strings]{underscore}
\usepackage{amsmath}  
\usepackage{listings}
\usepackage{amsthm}

\usepackage{amsfonts}  

\usepackage{hyperref}

\usepackage{hyphenat}
\hyphenation{block-chain}
\hyphenation{side-chain}

\newtheorem{theorem}{Theorem}

\journal{Computer Networks}

\begin{document}

\begin{frontmatter}



\title{Survey of Crosschain Communications Protocols}

\author[label1,label2]{Peter Robinson}
\address[label1]{ConsenSys Software R\&D}
\address[label2]{School of Information Technology and Electrical Engineering, University of Queensland, Australia}

\fi

\begin{abstract}
Crosschain communications allows information to be communicated between blockchains. Consensus in the context of crosschain communications relates to how participants on one blockchain are convinced of the state of a remote blockchain. It describes how parties associated with a source blockchain come to agreement and communicate with a destination blockchain such that information from the source blockchain can be trusted. This paper surveys crosschain communications protocols, presenting them based on the top-level usage scenarios they are trying to meet: value swapping, crosschain messaging, and blockchain pinning. It analyses how each protocol achieves crosschain consensus, what trust assumptions are made, their ability to operate successfully in permissionless and permissioned blockchains contexts, and whether the protocol delivers atomic updates across blockchains. 
\end{abstract}

\ifARXIV
\else

\begin{graphicalabstract}
\end{graphicalabstract}

\begin{highlights}
\item Analysis of how crosschain communication protocols achieve consensus.
\item Analyses the applicability of crosschain communications protocols to permissionless and permissioned blockchains.
\item Analysis of whether crosschain communications protocols deliver atomic updates across blockchains.
\item Comparison of crosschain communications protocols.

\end{highlights}


\end{frontmatter}


\fi

\section{Introduction}
\label{sec:introduction}
Crosschain Communications refers to the transferring of information between one or more blockchains. Crosschain communications is motivated by two requirements common in distributed systems: accessing data and accessing functionality which is available in other systems. The first requirement, accessing data in other systems, has been previously achieved by use of distributed query languages, for example SPARQL Federated Query 1.1~\cite{sparql-w3} and the Resource Description Framework 1.1~\cite{rdf-w3}. The second requirement, accessing functionality in other systems, has been achieved by use of technologies such as Remote Procedure Calls (RPC)~\cite{rpc}, Common Object Request Broker Architecture (CORBA)~\cite{corba}, and Representational State Transfer (REST)~\cite{rest}.

Whereas previous distributed systems have operated with implicit trust, blockchain systems operate in partially trusted or untrusted environments. That is, consumers of distributed query languages issue queries to single servers and implicitly trust the results. Similarly, entities issuing remote procedure calls assume that if the system the call is issued to indicates the call has been executed, then it has been executed. Additionally, systems that execute remote procedure calls often execute the calls based on little or no authentication. In contrast to this implicit trust model of previous distributed systems, blockchain systems are designed to be Byzantine Fault Tolerant (BFT). This means that blockchain systems can tolerate node failures, network failures, and malicious actors. Rather than relying on a single entity, blockchain system nodes come to consensus on the validity of proposed transactions. In the context of crosschain communications, this means that cross-blockchain systems need to be BFT, and not rely on single parties for trust.

Crosschain Consensus is the technique by which nodes or entities on a destination blockchain know that nodes or entities on a source blockchain have come to agreement on some fact. It allows information from a source blockchain to be trusted on a destination blockchain. Understanding how crosschain consensus is achieved and the underlying trust assumptions of the crosschain communications protocol is important when evaluating the appropriateness of a protocol for use with permissionless and permissioned blockchains.

Safety~\cite{safety-liveness}, Liveness~\cite{safety-liveness}, and Atomicity~\cite{reed1983} are properties of crosschain communication protocols. The safety property captures the notion that bad states are not reachable. The liveness property captures the notion that good states are eventually reached. For atomicity in the context of crosschain communications, the safety property translates to: when the protocol finishes, the state updates on all blockchains involved in the crosschain transaction are either committed or rolled back. The liveness property states that the atomic crosschain transaction protocol eventually (after finite amount of time) terminates. 

Multiple previous studies have analysed crosschain communications protocols. Vitalik's \textit{Chain Interoperability}~\cite{chain-interoperability-vitalik} (2016) and Johnson and Robinson's \textit{Sidechains and Interoperability}~\cite{johnson2019a} (2019) provide a general review of techniques that were available at their respective times of publication. Zamyatin et al.'s \textit{SoK: Communication Across Distributed Ledgers}~\cite{sok-comms-across} developed a framework to design new and evaluate existing crosschain protocols, focusing on the inherent trust assumptions thereof. Wang's \textit{SoK: Exploring Blockchains Interoperability}~\cite{sok-exploring-bc-interop} attempts to categorise crosschain solutions and determine advantages and disadvantages of each solution. Belchior et al.'s \textit{A Survey on Blockchain Interoperability: Past, Present, and Future Trends}~\cite{survey-blockchain-interop} explores the categories and sub-categories of blockchain interoperability systems. 

This paper is a survey of deployed crosschain communications products and other important crosschain communications protocols, as of June 2021. The goal has been to describe each protocol in enough depth such that readers can reason about the analysis of each protocol. As such, this paper references but does not describe some protocols that are similar to the protocols described in this paper. Additionally, this paper has likely missed some protocols as new protocols are being proposed and implemented every day. 

This paper categorises crosschain communications techniques according to what capabilities they aim to deliver. It analyses how the crosschain communications protocols to achieve crosschain consensus, and the underlying trust assumptions of the protocols. It analyses whether the consensus protocols have safety, liveness, and atomicity properties. Additionally, the properties of the protocols are analysed in terms of whether they are suitable for permissionless blockchains, permissioned blockchains, or both. Finally, protocols are classified according to whether they require changes to blockchain software to operate or act as a blockchain application. 

The remainder of this paper is divided into seven sections. Section~\ref{sec:scenarios} \textit{Scenarios} describes the main capabilities that crosschain communications protocols aim to deliver. Section~\ref{sec:attributes} \textit{Required Attributes} explains important properties of crosschain communications protocols and the properties that need to be maintained for the protocol to be relevant to permissionless and permissioned blockchains. Sections~\ref{sec:valueswaptechniques} to \ref{sec:pinning}, \textit{Value Swap Techniques}, \textit{Crosschain Messaging Techniques}, and \textit{Pinning Techniques} surveys the main crosschain communications protocols. This paper closes with a discussion about how the protocols compare in Section~\ref{sec:discussion} and future research directions in Section~\ref{sec:future}.

\section{Scenarios}
\label{sec:scenarios}
Crosschain Communications facilitate a variety of usage scenarios~\cite{eea-cross-usecases} including: Crosschain Decentralized Exchanges, Crosschain Asset Trading, Crosschain Decentralized Asset Transformation, Crosschain Backup, Private Chain as a Key Manager, Information Syndication, Logistics and Finance Blockchain Collaboration, National Data Sovereignty, and Oracle blockchains. These high-level usages are based on the core capabilities: atomic swaps, crosschain messaging, and pinning. The crosschain consensus protocols described later in this paper are categorised based on which of these the protocol primarily aims to delivery. 

\subsection{Atomic Swaps}
\label{sec:senarios:swaps}

\begin{figure}
\includegraphics[width=\linewidth]{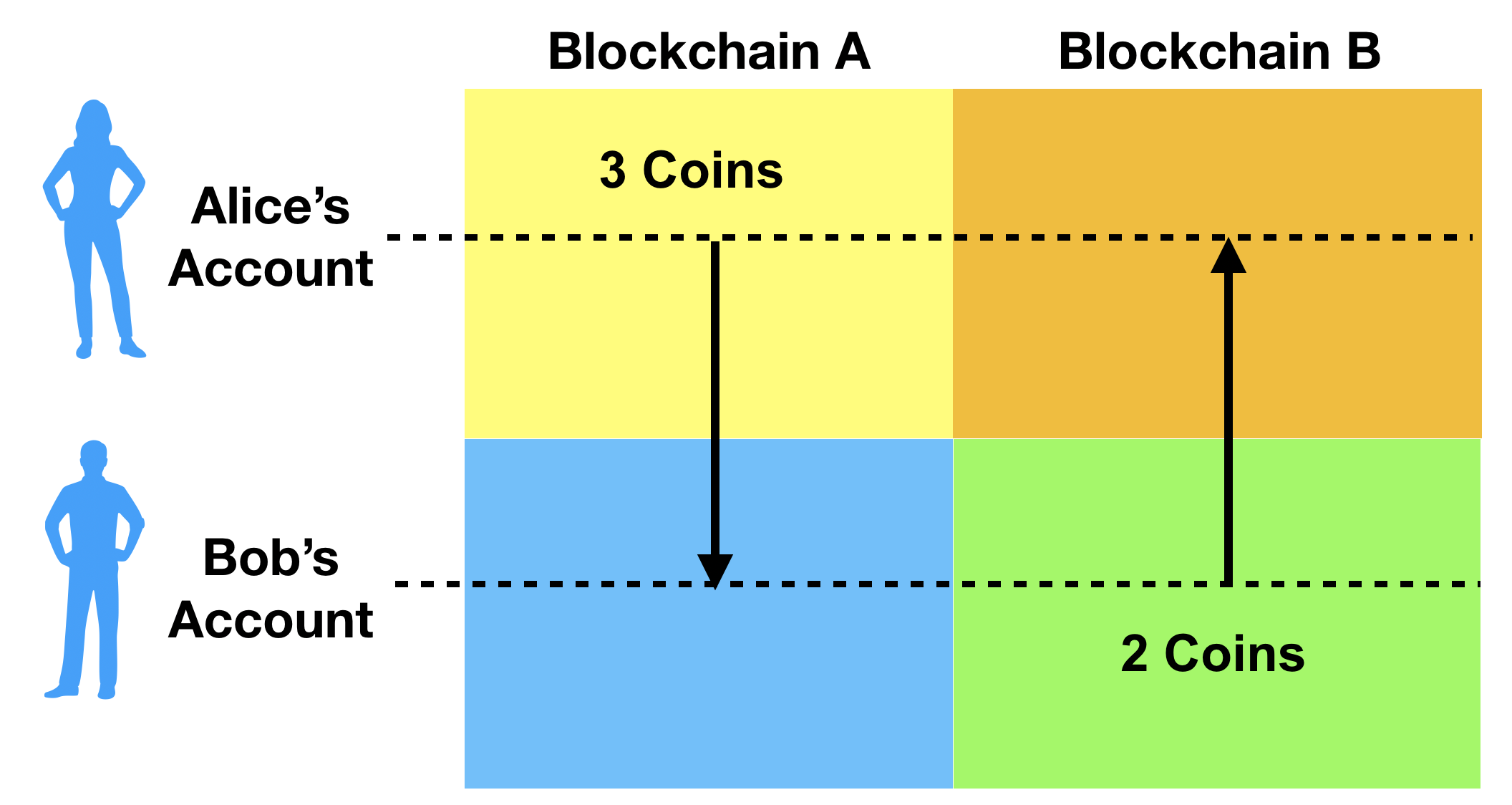}
\caption{Atomic Swap between blockchains}
\label{fig:atomic-swap}
\end{figure}

Users may have value on one blockchain and wish to move the value to another blockchain. A common way to affect a value transfer is to do an Atomic Swap. In Figure~\ref{fig:atomic-swap} Alice and Bob have accounts on both Blockchain A and B. Alice has three coins on Blockchain A. Bob has two coins on Blockchain B. Alice is prepared to swap her coins on Blockchain A for Bob's coins on Blockchain B. In an Atomic Swap, the coins are switched between Alice and Bob simultaneously. Importantly, if the transfer fails on either Blockchain A or B, then the transfer on the other blockchain fails too. 

Atomic swaps could be of the native currency of a blockchain such as Ether for Ethereum MainNet or Bitcoins for the Bitcoin blockchain. Alternatively, the value transfer could be for fungible tokens such as ERC 20 tokens~\cite{eip20, erc20-standard} or for non-fungible tokens such as ERC 721 tokens~\cite{eip721} which are represented as allocations within Ethereum smart contracts.

A special case of an atomic swap is when the value on the first blockchain is moved such that it is inaccessible and value is created on the second blockchain. In this situation, the value on the second blockchain needs to only be created if the value on the first blockchain can be proven to be inaccessible. This form of swap allows the value to be transferred across blockchains without the need for a counterparty to swap with. 

\subsection{Crosschain Messaging}
\label{sec:senarios:read}
\begin{figure}
\includegraphics[width=\linewidth]{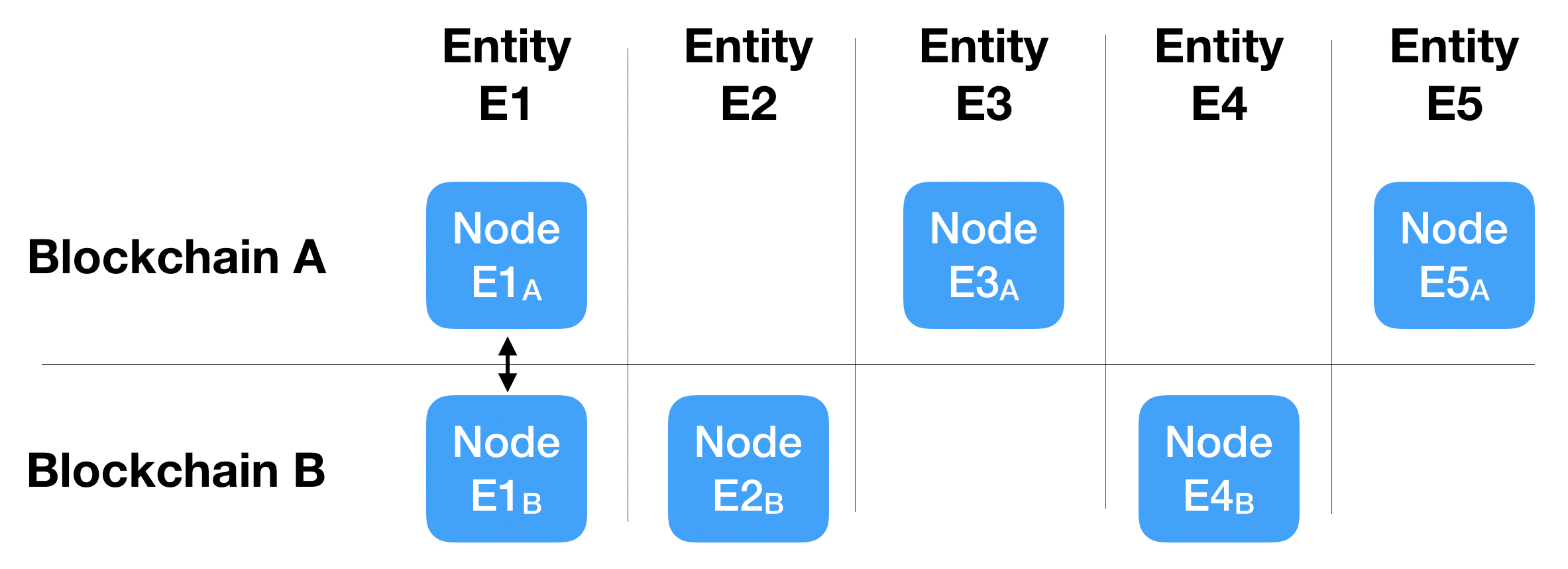}
\caption{Crosschain Messaging}
\label{fig:reading}
\end{figure}

Users may generate a message on one blockchain and consume the message on another blockchain. This allows values to be read and written across blockchains. Nodes on the source blockchain must convince nodes on the destination blockchain that the message from the source blockchain can be trusted. For example, consider the two blockchains in Figure~\ref{fig:reading}. Imagine that there is an application that executes a transaction on Blockchain A that generates a message that is consumed on Blockchain B. Nodes E2\textsubscript{B} and E4\textsubscript{B} must be convinced of the authenticity of the message despite having no access to Blockchain A. 

Atomic \index{atomic} crosschain transactions are transactions for which there is certainty that the updates across blockchains occur together. That is, the state updates associated with transactions are either committed on all blockchains or are discarded on all blockchains. In this way, the data across blockchains is guaranteed to have a consistent state.

\subsection{State Pinning}
\label{sec:senarios:pinning}
\label{sec:senarios:finalstatepinning}
\begin{figure}
\includegraphics[width=\linewidth]{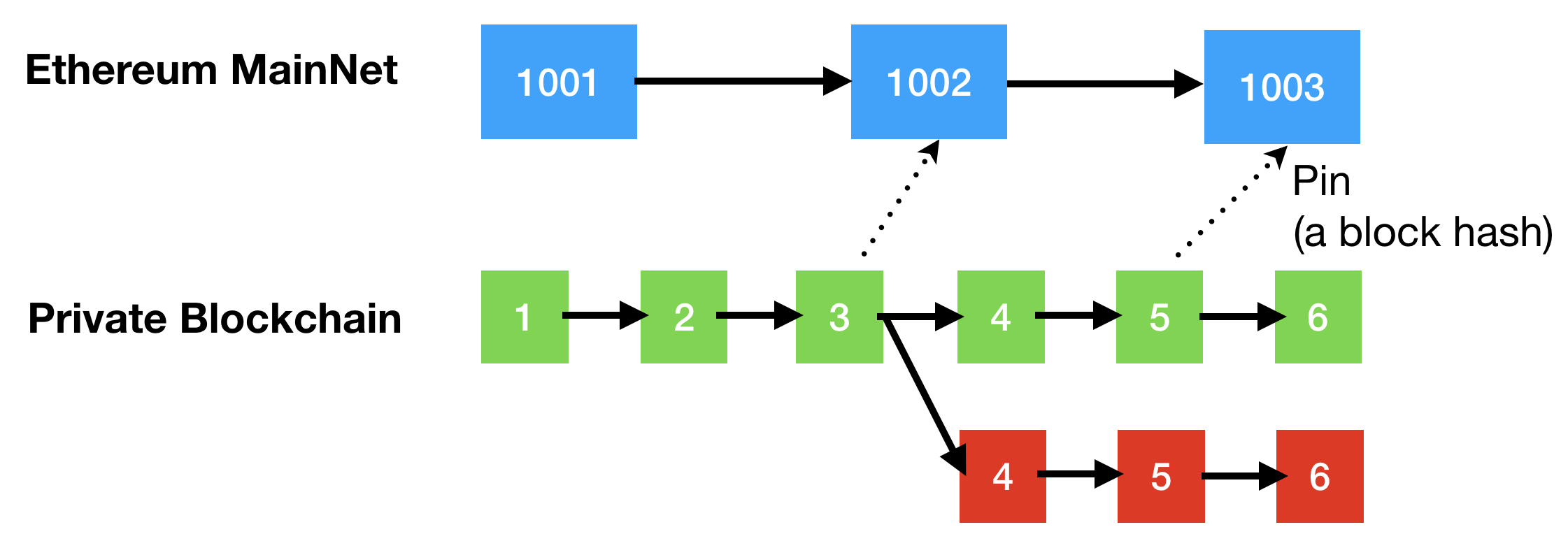}
\caption{Pinning state of private blockchain onto Ethereum MainNet}
\label{fig:pinning}
\end{figure}

State Pinning~\cite{anonpinning} is defined as including the state of one blockchain in another blockchain. For example, in Figure~\ref{fig:pinning} the block hash of a private blockchain is put into a smart contract on Ethereum MainNet. As the block hash from the private blockchain is included in Ethereum MainNet at a particular block number, it indicates that the state of the private blockchain can be represented by that Block Hash at that time. The block hash of a block is known as a ``Pin".

A majority of participants of the private blockchain could collude to alter the historical state of the chain~\cite{kaleido-relay}. For example, in Figure~\ref{fig:pinning}, the colluding participants could produce new blocks 4, 5, and 6 on the private blockchain. State Pinning allows minority participants of the chain to prove to governmental regulators and others that the state of the chain has been altered, by showing that the correct state matches the pinned Block Hash. That is, the minority participants could demonstrate the valid state of the blockchain at block 5 and the pin which has been included in Ethereum MainNet block 1003.

When the state of a private blockchain is pinned, it is important to maintain the privacy of the blockchain by not revealing the participants of the blockchain, ``Participant Privacy", or the blockchain block or transaction rate, ``Block Transaction Rate Privacy". Not disclosing the transaction rate of a private blockchain is important as attackers may be able to infer activity based on this. ``Anonymous State Pinning" hides the transaction rate of the pinned private blockchain and hides the identities of the participants, except in exceptional circumstances.

\begin{figure}
\includegraphics[width=\linewidth]{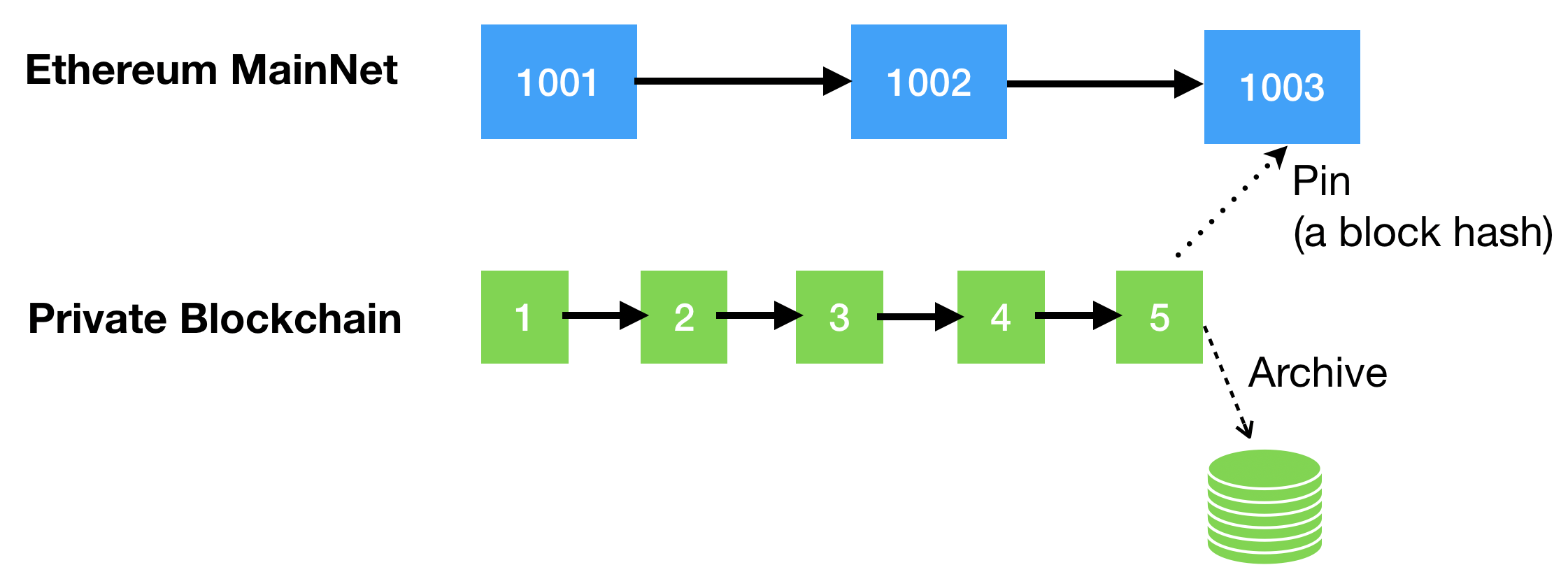}
\caption{Pinning state of private blockchain onto Ethereum MainNet}
\label{fig:finalstatepinning}
\end{figure}

A specialised use of pinning is to store the final state of a private blockchain prior to the blockchain being archived. Pinning the final state of a blockchain prior to archiving allows the blockchain to be reinstated later if needed at a state which can be verified. For example, in Figure~\ref{fig:finalstatepinning}, the private blockchain's last block, block 5, is pinned to Ethereum MainNet prior to the blockchain being archived. If at a later date the private blockchain needs to be reinstated, the reinstated state can be shown to be correct by comparing the block hash of the reinstated state with the pinned value.

\section{Required Attributes}
\label{sec:attributes}
Different types of blockchains and blockchain deployments have different crosschain communications system requirements. That is, in order to meet the goals of the blockchain deployment, the crosschain communications system may have specific requirements or limitations to meet such that they do not compromise the goals of the blockchain system that they are part of.

\subsection{Permissionless Blockchains}
Permissionless blockchains are blockchains that allow any node to join the network of nodes. Any account can submit a transaction on any node to the blockchain. Permissionless blockchains are commonly called \textit{Public Blockchains}\index{Public Blockchain} or \textit{Public Permissionless Blockchains}. Examples of public permissionless blockchains are Bitcoin and Ethereum MainNet.

Permissionless blockchains do not have any central control. They typically have many controlling nodes. Any node can join the network. Any node can create a block based on the consensus protocol rules of the blockchain.

Crosschain consensus protocols, like other consensus protocols, need a methodology to encourage good behaviour. \textit{Bad behaviour} could be due to malicious behaviour, but could also be due to network outages, system failures, misconfiguration, and software defects. The incentivisation scheme needs to match the type of blockchains that the crosschain protocol is being operated across. Permissionless blockchains rely on crypto-economic enforcement, such as charging transaction fees and slashing, to discourage bad behaviour~\cite{crypto-economics}.

\subsection{Permissioned Blockchains}
Permissioned blockchains typically restrict the nodes that can join the network to only authorised nodes. Additionally, permissioned blockchains may restrict which accounts are authorised to submit transactions. Permissioned blockchains are commonly called \textit{Private Permissioned Blockchains}, \textit{Private Blockchains} or \textit{Consortium Blockchains}. Examples of permissioned blockchains are blockchains that can be established by Enterprise Ethereum~\cite{enteth7} platforms such as Hyperledger Besu or ConsenSys Quorum, or other private blockchains platforms such as Hyperledger Fabric. Permissioned blockchains may allow any node to join the network but restrict the nodes that transactions can be submitted on. This type of network is typically known as a \textit{Public Permissioned Blockchain}.

Permissioned blockchains~\cite{enteth7} need to keep the list of participating nodes private. They need to keep the contents of transactions confidential. They need to keep the rate of transactions private. 

Permissioned blockchains are often between a small number of participants, By virtue of the small number of participants they have some level of centralisation. Some permissioned blockchains, such as Hyperledger Fabric and Corda have centralisation points~\cite{robinson2018a, hyperledger-fabric-doc, corda-doc}. Crosschain systems typically need to have no centralisation points if they are to be used with permissionless blockchains, whereas some level of centralisation may be acceptable for permissioned blockchains. 

Good behaviour is incentivised on permissioned blockchains by reputation and the use external enforcement such as taking legal action in a court of law~\cite{crypto-economics}.

\subsection{Safety, Liveness, and Atomicity}
As discussed in the Introduction (Section~\ref{sec:introduction}), safety~\cite{safety-liveness}, liveness~\cite{safety-liveness}, and atomicity~\cite{reed1983} are properties of crosschain communications protocols. The safety property captures the notion that bad states are not reachable. 

The liveness property captures the notion that good states are eventually reached. That is, the protocol finishes after a finite number of states. This property ensures assets across blockchains are not locked forever. Fischer et al.~\cite{distributed-consensus1985} showed that it is impossible to guarantee liveness for protocols that rely on asynchronous communications as anyone communicating party may take arbitrarily long time to respond. As such, the liveness property is only achievable if the underlying communications protocol is synchronous. 

Atomicity is the combination of safety and liveness properties. For atomicity in the context of crosschain communications, the safety property translates to: when the protocol finishes, the state updates on all blockchains involved in the crosschain transaction are either committed or rolled back. That is, the state updates on all blockchains will be consistent. The liveness property states that the atomic crosschain transaction protocol eventually, after finite amount of time, terminates. 

Herlihy et al.~\cite{crosschain-deals} defined liveness as, ``No asset belonging to a compliant party is escrowed forever", and safety as, ``every compliant party ends up with an acceptable payoff". Herlihy argues that actors in a blockchain system can be viewed as \textit{compliant} or \textit{deviating}, and that protocol analysis should not rely on threshold numbers of honest parties. This paper highlights the trust assumptions and, assuming those trust assumptions are met, analyses whether protocols have the safety and liveness properties.

\subsection{Crosschain Protocol Implementation Requirements}
Crosschain protocols can be classified as acting as applications or being part of blockchain platforms. When a crosschain protocol acts as an application, it does not need changes to the blockchain platform software to operate. It utilises contracts on the blockchain and servers external to the blockchain to operate. These protocols are appropriate for situations in which users are unwilling or unable to modify their blockchain platform software. In contrast, other protocols require changes to the underlying blockchain software to work. They are part of the blockchain platform. They do not need contracts or servers external to the blockchain to operate.


\section{Value Swaps Techniques}
\label{sec:techniques}
\label{sec:valueswaptechniques}

\subsection{Hashed Timelock Contracts}
\label{sec:techniques:hash-timelock-contracts}
Hashed Timelock Contracts (HTLC)~\cite{htlc-org,hashtimelock} are a mechanism for trustless crosschain atomic swaps. That is, the technique allows two parties to swap value on one blockchain for value on another blockchain. Agreement occurs off-chain, with on-chain consensus used to ratify the earlier off-chain agreement. Consensus is formed between the two parties on: who the two parties are (identified by account numbers that could be different on each blockchain), and on the amounts of value on each blockchain to be exchanged. HTLCs form the basis of the value swap mechanism in Poon and Dryja's Bitcoin Lightning Network~\cite{lightning2016} and the Dogecoin to Ethereum bridge~\cite{dogethereum}.

\begin{figure}
\includegraphics[width=\linewidth]{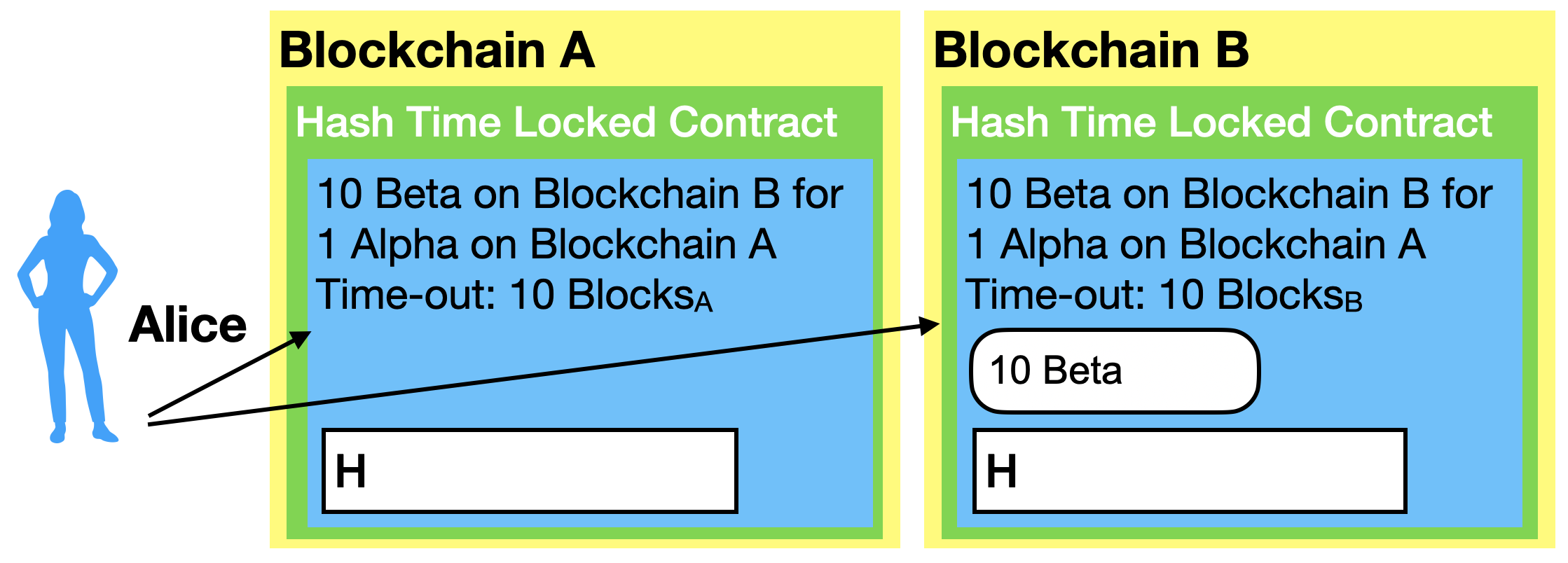}
\caption{Hash Timelock Contract Set-up}
\label{fig:htlc-setup}
\end{figure}

For example, Alice wishes to swap her ten Beta tokens on Blockchain B for Bob's one Alpha token on Blockchain A. To affect this swap, as shown in Figure~\ref{fig:htlc-setup}, Alice deploys a Smart Contract on the two blockchains. She generates a random number \texttt{R} and calculates a \textit{commitment} value \texttt{H} using Equation~\ref{equation:htlc1}. She submits the commitment value \texttt{H} and a timeout to both contracts. The timeout can be in terms of block numbers or wall clock time. She then deposits her ten Beta tokens into the contract on Blockchain B, configuring the contract such that Bob can withdraw the tokens by presenting the random number \texttt{R} or she can withdraw the tokens after the timeout.

\begin{equation}
\label{equation:htlc1}
H = Message Digest(R)
\end{equation}

\begin{figure}
\includegraphics[width=\linewidth]{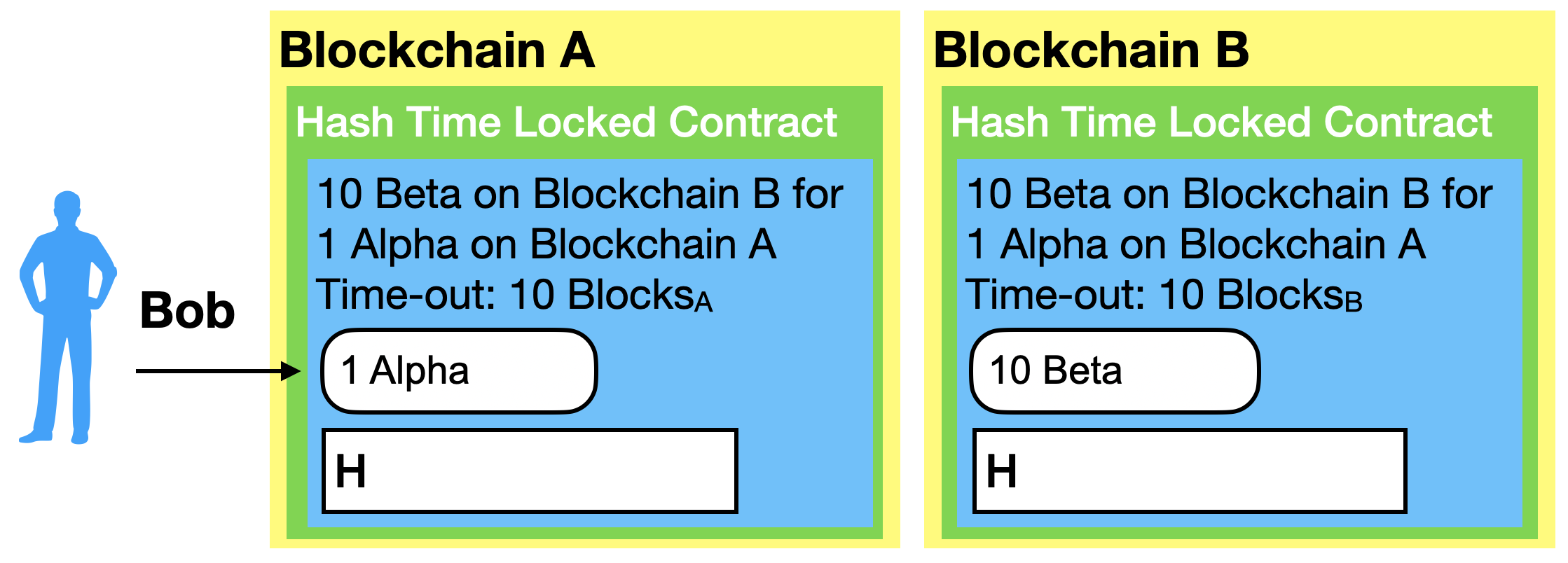}
\caption{Hash Timelock Contract Accept Offer}
\label{fig:htlc-accept}
\end{figure}

Next, as shown in Figure~\ref{fig:htlc-accept}, Bob deposits his one Alpha token into the contract on Blockchain A. This signals that Bob accepts the offer. Bob's Alpha token is now escrowed until either Alice submits the random number \texttt{R}, or the timeout expires. 

\begin{figure}
\includegraphics[width=\linewidth]{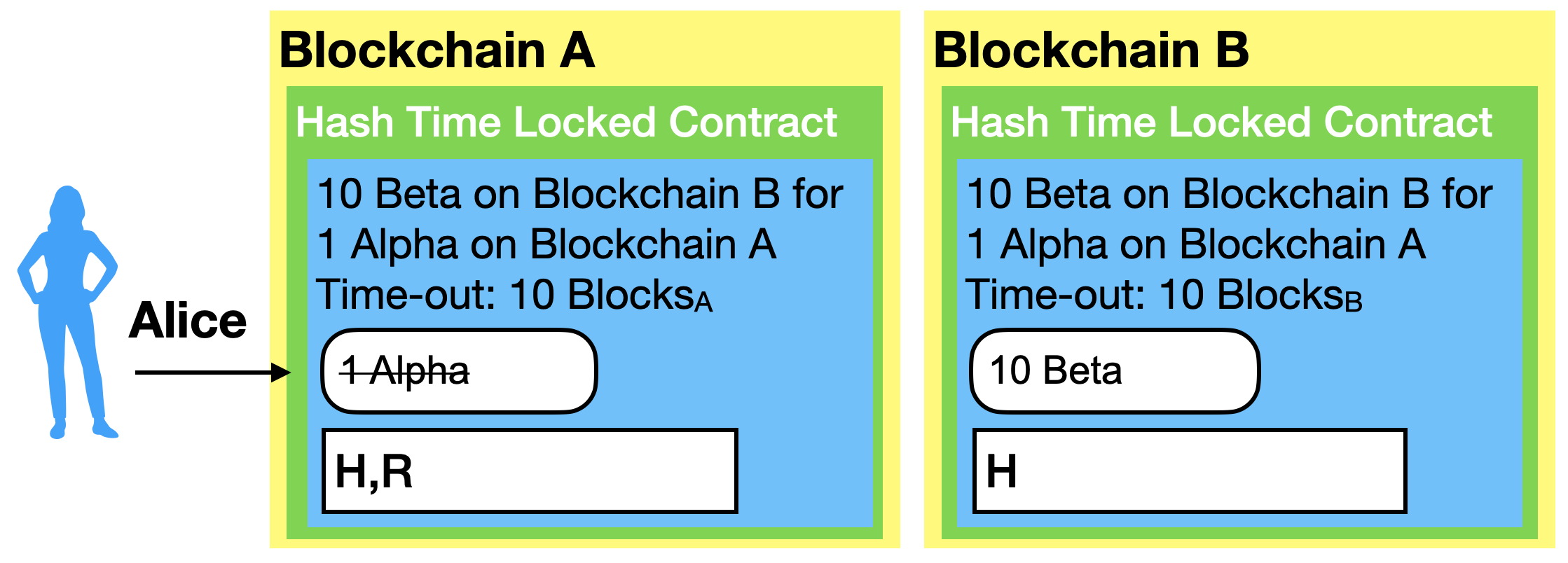}
\caption{Hash Timelock Contract Alice Withdrawal}
\label{fig:htlc-alice-finalise}
\end{figure}

As soon as Bob has deposited his token, Alice can submit the random value \texttt{R} to the contract on Blockchain A to withdraw the Alpha token, as shown in Figure~\ref{fig:htlc-alice-finalise}. That is, Alice can withdraw the token on Blockchain A if the smart contract on Blockchain A contains \texttt{H} and \texttt{R}, and Equation~\ref{equation:htlc1} holds for the values. 

\begin{figure}
\includegraphics[width=\linewidth]{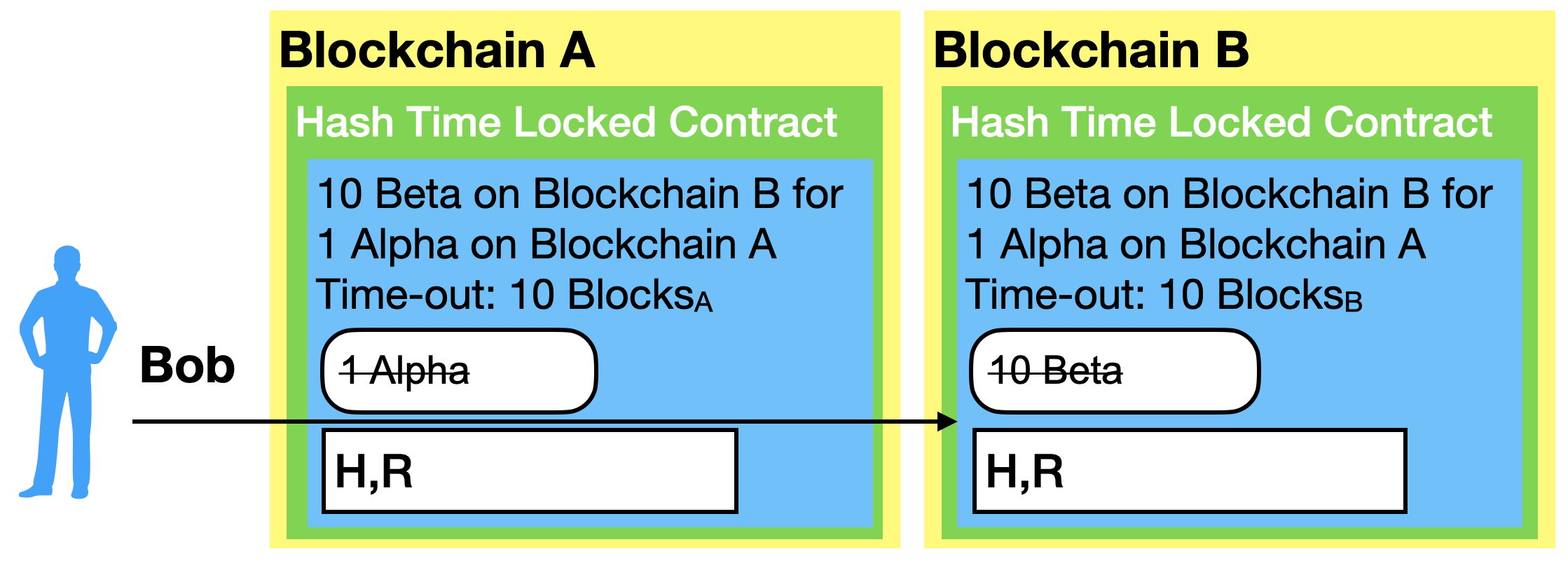}
\caption{Hash Timelock Contract Bob Withdrawal}
\label{fig:htlc-bob-finalise}
\end{figure}

As soon as Bob sees Alice's transaction in Blockchain A's transaction pool, he can copy random value \texttt{R} from the transaction. As shown in Figure~\ref{fig:htlc-bob-finalise}, he can then submit a transaction on Blockchain B containing the value to allow him to withdraw the Beta tokens from the contract in Blockchain B. That is, he can withdraw the Beta tokens if he presents a value whose commitment is \texttt{H}.

Alice and Bob can withdraw their tokens they have deposited if the timeout expires, and the correct random value \texttt{R} has not been submitted to the contracts. If the timeout is specified in blocks, one should consider that different blockchains generate blocks with different block periods. Even if the target block period on the two blockchains is the same, the timeout on the two blockchains cannot be precisely the same as block generation is not synchronised across blockchains. As such, blocks which reflect the timeout block number on each blockchain will be produced at different times.

If the blockchains both support the ability of checking wall clock time, then wall clock time could be used. For blockchains that support wall clock time, the time is likely to be the time at which the block containing the transaction accessing the time value was generated. In Ethereum, the node mining the block chooses the block's timestamp. Other nodes in the network will accept the block, as long as the block timestamp is within fifteen seconds of the current time. As such, miners in Ethereum and other similar blockchains have some limited influence as to which block a Hash Timelock Contract expires in.

The system can be subject to misuse~\cite{chain-interoperability-vitalik, hedging-against-sore-loser, htlc-game-theoretic}. Due to the price of Alpha tokens falling relative to the price of Beta tokens, Alice could determine that she should not complete the transfer. She could withhold random value \texttt{R}. If the token prices changed in the opposite direction, Bob could determine that he should not complete the transfer. In this situation he would not deposit his tokens. In this case Alice or Alice and Bob can withdraw their original tokens after the timeout. However, this \textit{griefing} attack (also known as a \textit{sore loser} attack) means that Alice or Alice and Bob do not have utility to redeploy their tokens during the timeout period. It should be noted that recent research indicates that rational participants will execute transfers or aborts immediately~\cite{htlc-game-theoretic}. 

\textit{Collateral Deposits}~\cite{hedging-against-sore-loser} have been proposed as a method of compensating parties in the case of \textit{griefing} attacks. Each party submits a small deposit to a contract. In the case of that a party withdraws from the protocol, the other party is able to withdraw the first party's deposit. Rather than setting up deposits for each value swap, Harz et al. have proposed using deposits across multiple value swaps~\cite{promise-deposits}.

HTLCs have the safety issue that Bob could lose his funds if he goes offline for longer than the timeout period prior to withdrawing his funds. He could try to submit his withdrawal transaction, but it may not be included in Blockchain B. An additional safety issue is that Alice could submit her transaction to withdraw the tokens on Blockchain A, thus revealing the random number \texttt{R} in Blockchain A's transaction pool, but the transaction might never be included in the blockchain. The reason the transactions might not be included could be due to transaction costs on the blockchains increasing, and hence the transaction being rejected. If the blockchains were private blockchains, and some validators did not want to process the transaction, they could delay the transaction being included in a block. Depending on the consensus protocol, this delay could be such that the transaction is not processed prior to the timeout. 

HTLCs do not suffer from liveness issues. If funds are not transferred prior to the timeout, then the participants are able to withdraw the funds. Even in the case of the safety issue described above, where a participant goes offline, the contract is not blocked from use once the timeout has expired.  

As HTLCs allow trustless exchange of assets on two separate blockchains, they are suitable for Permissionless Blockchains. Using HTLCs, a trustless exchange mechanism, on a permissioned blockchain, a semi-trusted blockchain system could appear to be more than is needed. That is, it could be conjectured that simply forwarding the exchange request to the semi-trusted blockchain would be enough. However, simplistic trusted exchange mechanisms like this do not provide atomic behaviour. Given HTLCs provide atomic behaviour, their extra complexity over simplistic trusted exchanges is warranted for permissioned blockchains.

\subsection{Interledger's STREAM Protocol}
\label{sec:interledger}
\begin{figure}
\includegraphics[width=\linewidth]{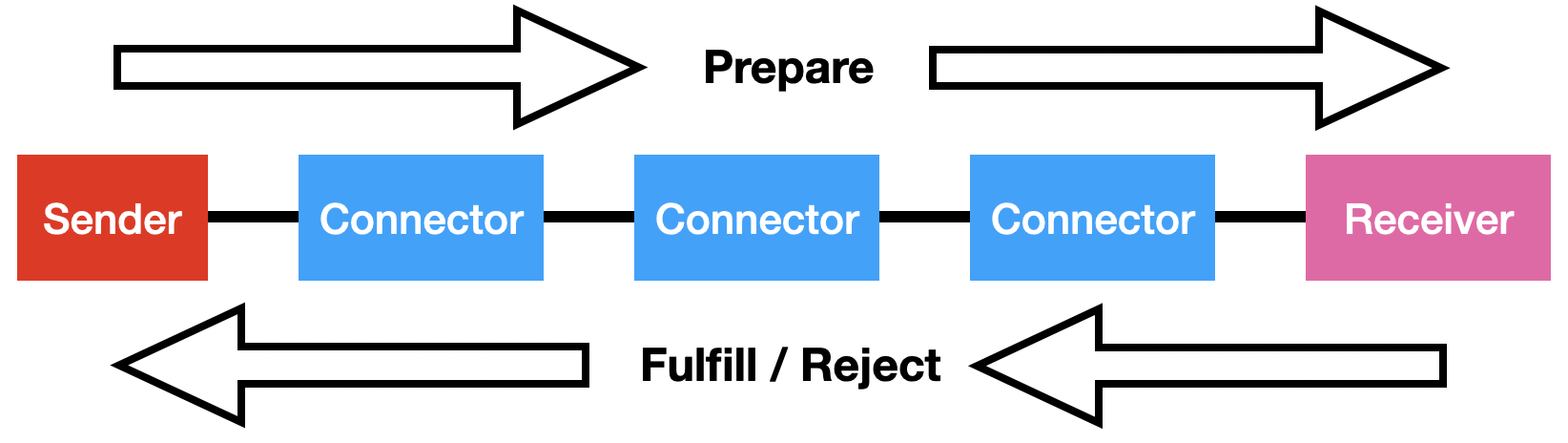}
\caption{Interledger Protocol}
\label{fig:interledger}
\end{figure}

Interledger~\cite{interledger2021} is a set of payment protocols for sending value across heterogeneous blockchain systems. The core of the protocol is Interledger's transport protocol ILPv4. As shown in Figure~\ref{fig:interledger}, Senders communicate via Connector nodes to Receivers. Senders send request messages called \textit{Prepare} packets and Receivers respond with response messages called \textit{Fulfill} packets or error messages called \textit{Reject} packets. 

Consensus is achieved by the Prepare messages containing a hash and the Fulfill messages containing the corresponding preimage. The Prepare message provides a commitment by the Sender to pay. The Fulfill message provides acknowledgment by the Receiver that they have received the funds. None of the Connectors knows the preimage, and hence can not forge the payment receipt. Given the use of hash-locking, this is a trustless communications methodology.

Interledger's STREAM protocol~\cite{interledger2021-stream} is a bi-directional, multi-stream, packetized payment and data protocol built on top of ILPv4, based on the ideas of packetized payments~\cite{packet-payments}. Large payments are split into economically insignificant pieces. Each piece is sent in a separate \textit{Prepare} message. Receivers send acknowledge \textit{Fulfill} messages indicating the total amount received. The idea is that untrusted counterparties can exchange economically significant amounts by exchanging many economically insignificant payments. 

An attacker could attempt to execute a griefing attack on the packtized payment protocol in a similar way to how the attacks are performed on HTLCs. By sending small economically insignificant amounts in the packetized payment protocol, such griefing attacks only affect small amounts of value. Dubovitskaya et al.~\cite{packet-payments-analysis} determined that without collateral deposits users may abort an exchange if there are price fluctuations mid-transfer.

\subsection{BTC Relay: Simplified Payment Verification}
\label{sec:techniques:transfer-block-headers}
\label{sec:btcrelay}
\begin{figure}
\includegraphics[width=\linewidth]{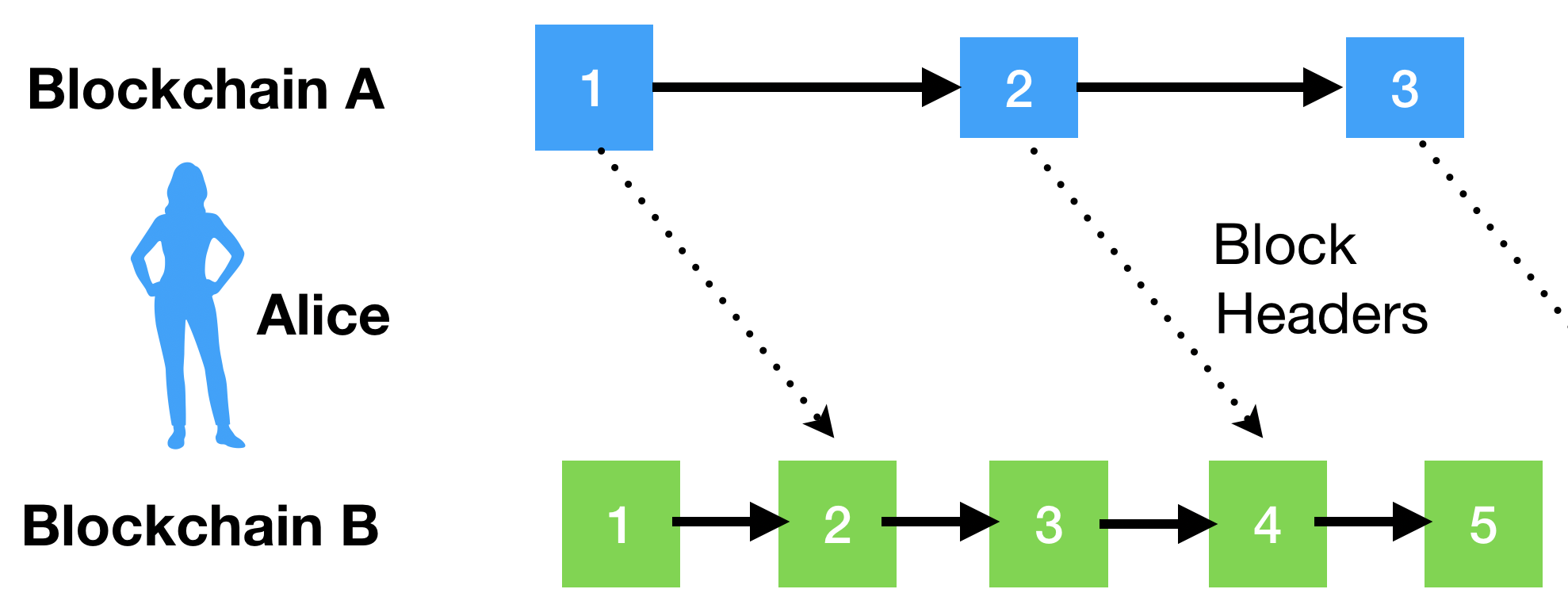}
\caption{Block Header Transfer}
\label{fig:transfer-block-header}
\end{figure}

\begin{figure}
\includegraphics[width=\linewidth]{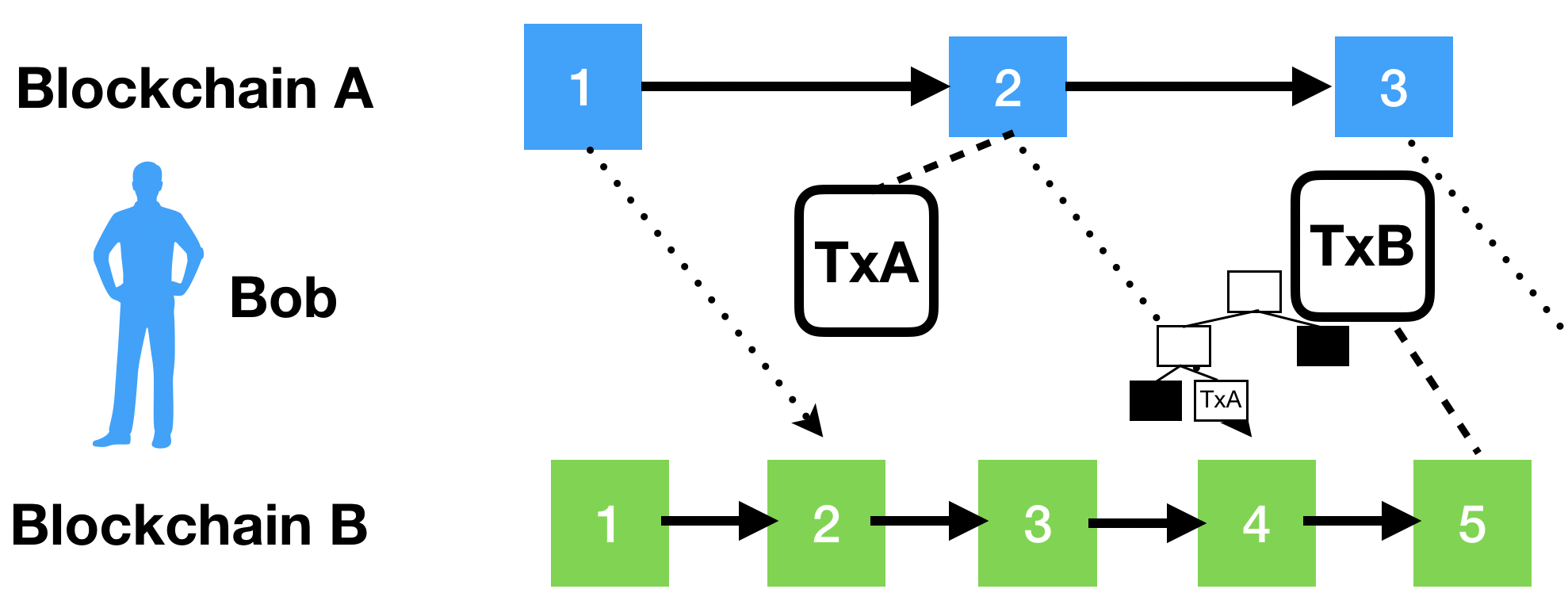}
\caption{Transaction Transfer}
\label{fig:transfer-transaction}
\end{figure}

BTC Relay~\cite{btc-relay} allows users of Ethereum MainNet to do actions based on Bitcoin transactions using the Simplified Payment Verification (SPV) approach described in the Bitcoin White Paper~\cite{nakamoto2008}. SPV relies on Block Headers being transferred between blockchains. For example, in Figure~\ref{fig:transfer-block-header} Alice observes blocks being produced on Blockchain A. For each block produced on Blockchain A, she submits the block header to Blockchain B by submitting a transaction to a contract. Anyone, in this example Bob in Figure~\ref{fig:transfer-transaction}, can then submit a transaction to the contract on Blockchain B that includes the block number, the transaction which is said to have occurred on Blockchain A, and a Merkle Proof that proves that the transaction occurred on Blockchain A in a certain block. The contract calculates the hash of the transaction said to have occurred on Blockchain A and calculates the Root Hash using the partial Merkle Tree. If the calculated Root Hash matches the value in the block header indicated by the block number, then the transaction will be deemed to have occurred on Blockchain A in the specified block.

The safety of an SPV based system relies on the block headers being transferred such that they can be trusted. The liveness of the system depends on the block headers actually being transferred. If the entity(ies) transferring block headers stops transferring them, then no transactions can be confirmed on the second blockchain.

A typical action in BTC Relay is to move Ether from one account to another account based on a Bitcoin transaction transferring BTC from one account to another account. Entities called \textit{Relayers} are compensated for posting Bitcoin block headers to a Smart Contract on Ethereum MainNet. BTC Relay relies on Proof of Work (PoW) mining difficulty for its security. Multiple active Relay nodes must be prepared to post the block header for each block. In this way, if one Relay node posts a block header of a fork of the chain, other Relay nodes can post the block header of the longest chain. Transactions can only be validated if the block header they relate to is on the longest chain and if at least six block headers have been posted on top of the block header that the transaction relates to~\cite{btc-relay-source}. As attackers can not produce a longer chain than the main Bitcoin blockchain due to the mining difficulty, they are unable to confirm transactions based on a malicious fork. 

Users of BTC Relay pay a fee to confirm Bitcoin transactions on Ethereum. This fee is used to compensate Relayers. However, if no users are confirming transactions on BTC Relay, then Relayers are not compensated for submitting blocks to the BTC Relay contract on Ethereum. This means that the Relayers are not incentivised to submit Bitcoin block headers. If no block headers are submitted, then transactions can not be confirmed. This liveness issue has occurred in the production deployment of BTC Relay as no Bitcoin block headers have been transferred to BTC Relay's contract on Ethereum MainNet since February 2018~\cite{btc-relay-etherscan}. To allow recent Bitcoin transfers to be accessible on Ethereum MainNet, Bitcoin block headers could be submitted to the contract so that it was up to date. Alternatively, a new BTC Relay contract could be deployed, with a recent Bitcoin block header used as the genesis block header. 

A detailed security analysis of BTC Relay was conducted in 2016 by Martin Swende~\cite{btc-relay-hacks}. This analysis identified implementation issues such as being able to avoid fees while confirming Bitcoin transactions, but no safety issues relating to crosschain consensus. 

Lys et al.~\cite{rswap} analysed the application of BTC Relay to value swapping and determined safety and liveness issues. Bob could lock value in a contract on Ethereum that could be unlocked by a confirmed transfer on Bitcoin. If Alice and Carol both submitted transactions on Bitcoin to transfer money to Bob, only one of them would be able to redeem their Bitcoin transaction. The other would be unable to get a refund on Bitcoin and would not be able to use the confirmed Bitcoin transaction to access the locked value on Ethereum. The liveness issue arises as Bob could lock his value in the contract on Ethereum, but then no one could submit a transaction on Bitcoin, thus locking his value in the contract on Ethereum forever. 

BTC Relay relies on the source blockchain using PoW consensus algorithm that can be verified on the destination blockchain. That is, Ethereum Smart Contracts can be created to verify Bitcoin's PoW algorithm. However, programmatically verifying Ethereum's time and memory hard Ethash PoW algorithm requires large amounts of memory and hence is difficult to execute in smart contract languages. An additional limitation is that the hashing power of the source blockchain must be large enough such that attackers can not create forks and affect transfers based on the forks. This effectively limits BTC Relay's approach to having Bitcoin as the source blockchain.

\subsection{XCLAIM}
\label{sec:xclaim}
The XCLAIM~\cite{xclaim} protocol uses the BTC Relay protocol to send transaction inclusion proofs from Bitcoin to Ethereum. Building on this underlying crosschain consensus, the protocol allows for atomic transfers of value from Bitcoin to Ethereum and back. The protocol uses an off-chain entity called a \textit{Vault} and an \textit{Issuing Smart Contract} (iSC) to facilitate transfers. 

When the Bitcoins are transferred to the Ethereum blockchain, they are sent to the Vault's address on the Bitcoin blockchain and \textit{Wrapped Bitcoin} are issued on the Ethereum blockchain. To transfer Bitcoins to Ethereum a \textit{Requester} commits to the transfer by locking up a small amount of Ether in an \textit{Issuing Smart Contract} (iSC) on Ethereum and specifies an Ethereum address that the Wrapped Bitcoin should be issued to. The iSC locks the corresponding amount of the \textit{Vault}'s Wrapped Bitcoin for a predefined delay. The Requester sends the appropriate amount of Bitcoin to the \textit{Vault} user's address on the Bitcoin blockchain. The Vault submits a transaction inclusion proof to the iSC, utilising BTC Relay. The iSC creates the appropriate amount of Wrapped Bitcoin and transfers the Wrapped Bitcoin to the Requester's address on Ethereum. If the Requester does not transfer the Bitcoin, then they lose their collateral Ether. If the Vault does not submit the inclusion proof transaction on Ethereum after a period of time, then the Requester can submit the transaction themselves, and the Vault is fined. 

To transfer Wrapped Bitcoin from Ethereum to Bitcoin, the Requester submits a transaction to call the iSC's \textit{Burn} function. This results in an Ethereum Event being emitted that indicates that Bitcoin should be unlocked on the Bitcoin blockchain. The Vault witnesses the event and, after waiting for the event to become probably final, transfers the appropriate number of Bitcoin to the Requester on the Bitcoin blockchain. They then use BTC Relay to prove that the transaction occurred on Bitcoin and then burn the Wrapped Bitcoin. If the Vault does not prove the transfer has occurred on Bitcoin quickly enough, the iSC fines them and reimburses the Requester.   

As XCLAIM relies on BTC Relay, it is limited to having Bitcoin as the source blockchain in the protocol. Bitcoin block headers have not been transferred to BTC Relay's contract on Ethereum MainNet since February 2018~\cite{btc-relay-etherscan}. As such, the underlying crosschain consensus required for this protocol when operating between Bitcoin and Ethereum MainNet is not operational. However, XCLAIM has been implemented on other blockchains such as Polkadot~\cite{interlay-interbtc}. The underlying BTC Relay protocol technology has been replicated on these other blockchains. 

\subsection{Pegged Sidechains}
\label{sec:pegged}
Pegged Sidechains~\cite{pegs2014}\index{Pegged Sidechains} propose Bitcoins be transferred between the Bitcoin blockchain and sidechains, in a similar way to SPV. The rationale for such sidechains was to allow for increased transaction rate and experimentation. The solution relies on publishing a Merkle Proof that a transaction to transfer Bitcoin was included in a block in the source blockchain and publishing the block headers that were produced based on that block, in the \textit{destination} blockchain. The authors of the paper~\cite{pegs2014} recommend users should consider providing 24 hours of block headers. If the hashing power of the source blockchain is significant, then it would be impossible for an attacker to produce forged blocks. Wood~\cite{polkadot2016} contends that the sidechain hashing power is unlikely to be sufficient to ensure security. Consequently, Bitcoins can be securely transferred to the sidechain from the Bitcoin blockchain, but not back. As such, this methodology should be viewed as having safety issues. Given the user affecting the transfer submits their own block headers, there is no requirement to wait for or incentivise relaying nodes to transfer block headers as there was in BTC Relay.  As such, this methodology does not suffer from liveness issues.

Pegged Sidechains are not generally applicable as they rely on the source blockchain using PoW consensus algorithm, and the hashing power of the source blockchain being large enough such that attackers can not create forks and affect transfers based on the forks. Additionally, PoW is not an appropriate consensus algorithm for private blockchains as organisations do not wish to allocate resources to mining of blocks~\cite{enteth7}. As such, they are inappropriate for private blockchains.

\subsection{Plasma}
\label{sec:plasma}
Minimum Viable Plasma~\cite{plasma-mvp} builds on the concept of Plasma's~\cite{poon2017} delegate Ethereum blockchains. Plasma chain operators create a Plasma Smart Contract on Ethereum MainNet and hold value deposited in the contract on a separate Plasma chain as Unspent Transaction Output (UTXO)~\cite{uxto} values in a binary Merkle tree ordered by transaction index. Transactions on the Plasma chain involve proving that an unspent output had not previously been spent. Plasma Cash~\cite{plasma-cash}, builds on the Minimum Viable Plasma approach to allow for the exchange of non-fungible assets. Each token has an identifier that represents the token's location in a sparse Merkle Tree. To spend a block, a proof needs to be submitted showing when the token has been used. 

For each Plasma variant, information is created on the Plasma chain by a chain operator after a deposit into the Plasma Smart Contract on Ethereum MainNet. To exit the Plasma chain, the owner of a token needs to submit a proof to the Plasma Smart Contract, proving that they own the token. Another user could challenge the \textit{exit} by submitting a fraud proof showing that a subsequent transaction occurred that changed the ownership of the token after the proof submitted by the user purporting to be the token owner. The challenge period is typically at least seven days. 

A drawback of Plasma is that users need to observe the Plasma Smart Contract continuously, checking for fraudulent exit proofs being submitted, to prevent tokens they rightfully own being converted into Ether. If a user is offline for the entire challenge period, then a nefarious actor could steal their tokens. The proofs are likely to be large. As such, exiting and challenging an exit on Ethereum MainNet could be costly. Another major issue is that Ethereum MainNet does not have sufficient capacity to allow for a \textit{mass exit} scenario, in which all members of a Plasma chain choose to exit the chain at the same time. 

From a crosschain consensus perspective, consensus is provided by users submitting exit proofs and fraud proofs. Plasma does not provide safety as nefarious actors could steal a user's tokens if a user is offline during a challenge period or if the nefarious actor could orchestrate a mass exit, thus preventing users from submitting fraud proofs.

\subsection{Wanchain}
\label{sec:wanchain}
Wanchain is a company that operates a public blockchain called \textit{Wanchain}~\cite{wanchain}. Wanchain is a fork of Ethereum that is designed to facilitate crosschain value transfers. Wanchain typically operates as a relay chain, with transfers from source chains to destination chains going via Wanchain. However, transfers can also go directly from one blockchain to another using Wanchain's crosschain techniques. The transfer technique used is the same between a source chain to Wanchain, Wanchain to a destination chain, or direct between two chains.

To facilitate a transfer, users deposit value into a source contract on a source blockchain, transferring ownership to the Storeman Group. Then at least 17 Storemen of a total 25 Storemen in a Storeman Group sign a transaction for the destination blockchain that causes value to be minted, with the owner being the user. The user can then withdraw the value from the destination contract on the destination blockchain. 

Each Storeman must stake a bond on Wanchain. The membership of the Storeman Group is changed each month; drawing members from a pool of possible Storemen based on who is prepared to stake the most. Storemen who are shown to have signed invalid transactions are slashed. Storemen are incentivised to operate the bridge correctly by being paid a fee at the end of each month.

The amount of value being transferred across a bridge is limited to less than the amount being staked to secure the bridge. This ensures Storeman Groups are incentivised to not commit fraud as the amount that can be slashed exceeds the amount that can be transferred.

The Wanchain crosschain system uses a threshold number of signers for crosschain consensus. Honest operation is encouraged by slashing and limiting the amount that can be transferred across the bridge to the staked amount. The system is not atomic as there is no guarantee that transactions on the destination blockchain will be executed correctly. In this case, the state across blockchains would be inconsistent.

\section{Crosschain Messaging Techniques}

\subsection{Ion Project}
\label{sec:ion}
The Clearmatics Ion\index{Ion} project uses block header transferring in the context of permissioned blockchains. It provides a framework and tools to develop crosschain smart contracts so that they execute if a state transition has occurred on another  blockchain~\cite{Clearmatics2018c,johnson2019a}. The system works by having a set of Relayers that wish to transfer block headers from one blockchain to another. At least a threshold number of the Relayers need to sign each block header for the block headers to be accepted by the receiving blockchain. Relayers only transfer a block header once the blocks they relate to are deemed to be final. For the scenario when the source blockchain is Ethereum, a user executes a transaction on the source blockchain that emits an event. The transaction's receipt contains the transaction hash, the block number, transaction number, and includes a list of event information. The event information is fed into the transaction hash calculation. The user can now execute a transaction on the destination blockchain, passing in as parameters information relating to the source blockchain's transaction: the block number, the Merkle Proof showing the transaction belongs to the block, and the event information. The code on the destination blockchain executes code that verifies the source blockchain's transaction information. If that information is found to be correct, then code can use the information in the event to perform some action. 

Ion allows for forwarding of events from one chain to another. This allows reading of information. It could be used for function call forwarding and messaging. However, the system does not facilitate atomic behaviour. There is no guarantee that an event will be forwarded. Even if the event is forwarded, there is no guarantee that transactions will be processed correctly and higher-level protocols will be able to operate correctly based on the event.

The block headers are submitted by Relayers separately. This means that there are multiple transactions on the destination blockchain per source blockchain block. Unless the Relayers want to sign and forward every block header, they need to analyse which blocks contain transactions that emitted events that are destined for the destination blockchain, and only forward those block headers.

Ion does not have any liveness issues assuming a threshold number of Relayers sign and submit block headers to the destination blockchain.

\subsection{ChainBridge}
\label{sec:chainbridge}
ChainBridge~\cite{chainbridge} is a technology to allow functions on one blockchain to call functions on another blockchain. Higher-level protocols have been created on top of the technology to provide ERC20 value transfer and ERC721 Non-Fungible Token transfer.

ChainBridge works by having applications call a \textit{deposit} function on a \textit{Handler} contract. The destination blockchain, contract address, function and parameter values are emitted in a \textit{deposit} event. Deposit events on the source chain are detected by a trusted set of off-chain \textit{Relayers} who wait for the block containing the transaction that emitted the event to become final. Each relayer submits a transaction containing the event to the destination blockchain to vote on the validity of the event. Once a threshold number of relayers have submitted the event, a handle function is called, which in turn calls the function on the contract that was contained inside the event.

The system has liveness issues as there is no certainty that a threshold number of relayers will submit an event. Additionally, there are safety issues as even if the threshold is reached, and the function is called on the destination blockchain, there is no certainty that the function will execute as expected. For example, the function may execute correctly on the source blockchain, updating state, and then the function on the destination contract may cause state to be reverted. In this situation, the source blockchain would have been updated but not the state on the destination blockchain.

\subsection{Celo Optics}
\label{sec:celo}
Celo Blockchain is a fork of Ethereum that uses Proof of Stake (PoS) consensus algorithm~\cite{celo-blockchain}. Celo Optics~\cite{celo-optics} is a technology to allow messages to be passed between blockchains. With Celo Optics, transactions on a source chain call a function on the \textit{Home} contract to create messages. Messages are accumulated into a Merkle Tree of messages. The messages form the leaves of the tree. Adding a message changes the root of the tree. A sequence of messages corresponds to a queue of roots. The root of the tree is signed by an off-chain \textit{Updater}. The signed root is submitted to the Home contract on the source blockchain. Off-chain \textit{Relayers} forward signed tree roots to \textit{Replica} contracts on destination blockchains. They could forward one root per message, or at some other frequency. The signed tree roots are only usable after a \textit{fraud proof window}. Off-chain \textit{Watchers} check that the off-chain Updaters are signing the correct tree roots, and that correct tree roots are being submitted to Replica contracts. If a Watcher detects an invalid tree root has been posted to a Replica contract, it can post the signed root to the Home contract, causing the Updater that signed the invalid root to be slashed. Off-chain \textit{Processors} hold all messages, and submit them, along with Merkle Proofs, to recipients on destination blockchains. Applications submit the transactions on the source blockchains to generate the messages and are the recipients on the destination blockchains.

Celo requires any function call or value swap protocol to be built on top of the underlying message passing scheme. There are no safety guarantees with the protocol. There is no certainty that messages will be delivered. Additionally, there is no certainty that once messages are delivered, that they will be able to be processed correctly. The latency depends on how often signed roots are communicated and how long the fraud window is. 

A key drawback of the system is that applications need to monitor for invalid messages and tree roots being posted by \textit{Relayers} and \textit{Processors}. They need to have logic in application code on the destination blockchain to ignore messages that are deemed invalid. This means that applications need to be involved in monitoring the operation of the underlying blockchain transfer protocol. This ability of an application to selectively ignore messages from remote blockchains means that the system operates in a permissioned environment. However, the requirement for slashing indicates that the system needs to operate in a public permissionless environment.

\subsection{Cosmos}
\label{sec:relay-chains}
\label{sec:cosmos}
\begin{figure}
\begin{center}
\includegraphics[width=.8\linewidth]{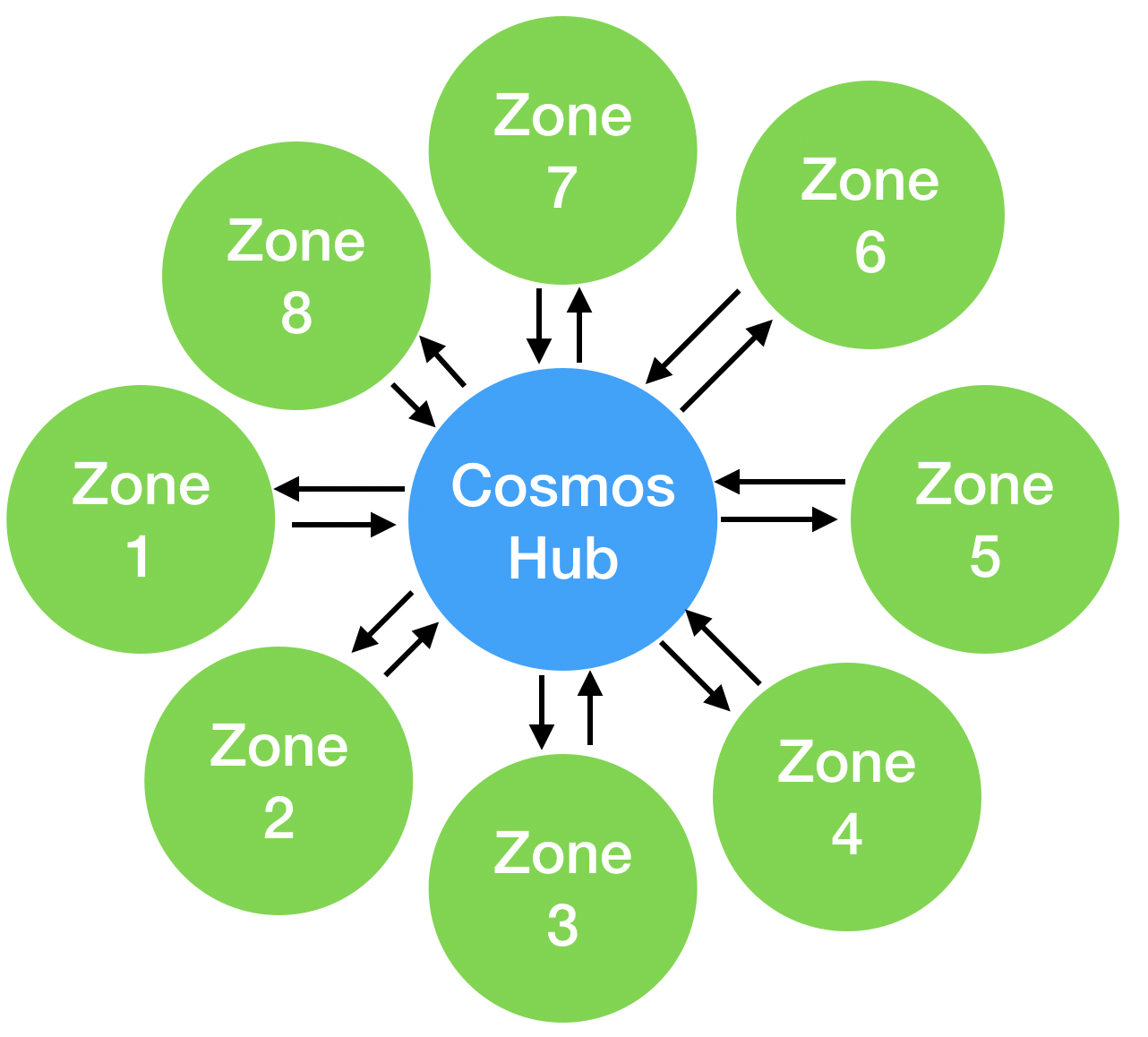}
\caption{Cosmos system of blockchains}
\label{fig:cosmos}
\end{center}
\end{figure}

Cosmos~\cite{cosmos2016} is a multi-blockchain system in which blockchains called Zones communicate transactions via a central blockchain called a Hub, as shown in Figure~\ref{fig:cosmos}. The Zones and the Hub typically use Tendermint~\cite{tendermint2018} a type of Practical Byzantine Fault Tolerance~\cite{pbft1999, pbft2002} consensus algorithm. The system is envisaged to allow heterogeneous blockchain communications, allowing the Zone blockchains to be permissioned or permissionless, to use alternative consensus algorithms, including algorithms that offer probabilistic finality, and allowing for completely different blockchain paradigms. 

Each Zone blockchain must have a set of validators. The Zone blockchain validators must trust validators in the Hub blockchain and visa-versa. 

\begin{figure}
\includegraphics[width=\linewidth]{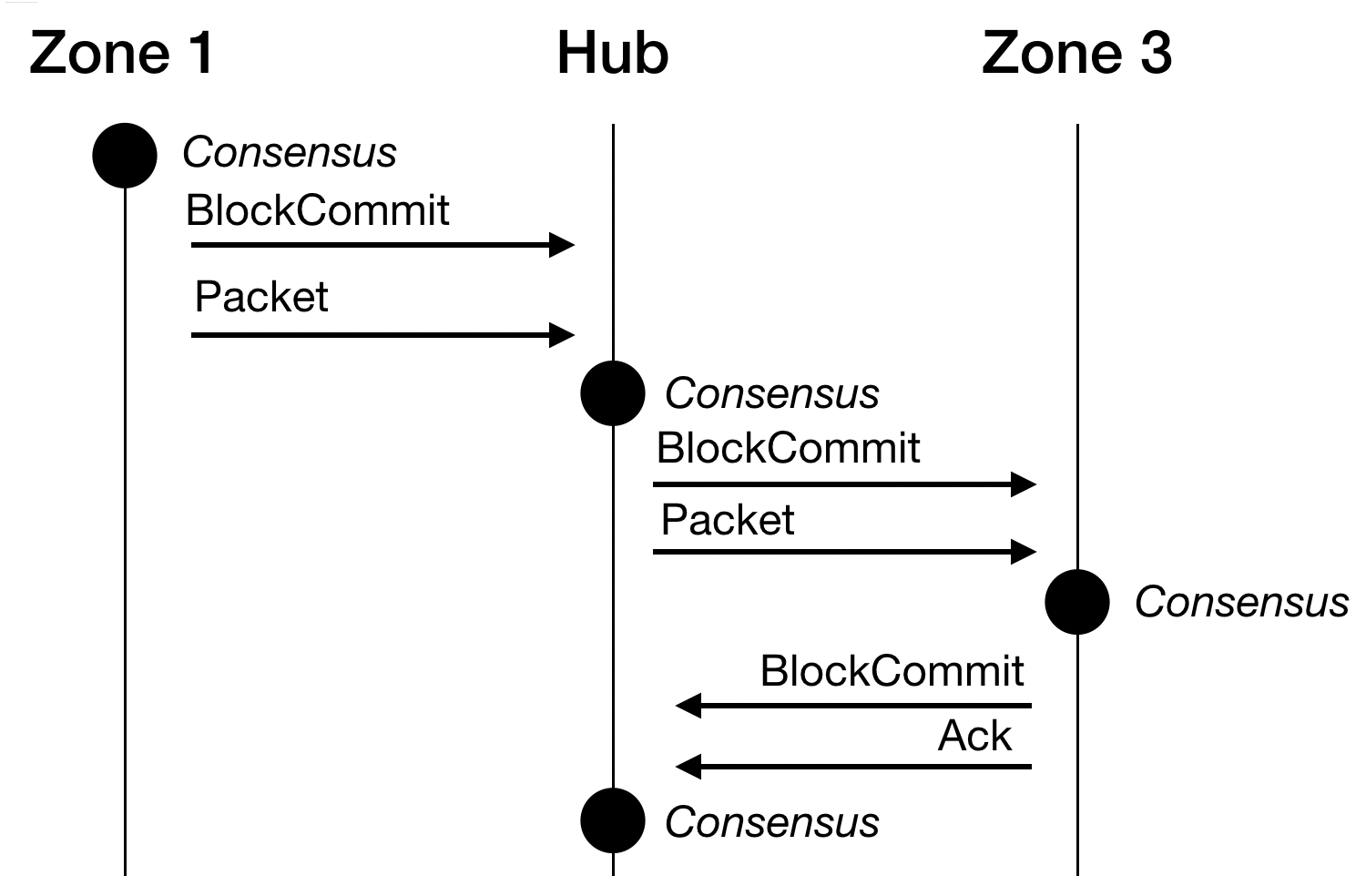}
\caption{Cosmos Inter-Blockchain Communications (IBC)}
\label{fig:cosmos-ibc}
\end{figure}

Figure~\ref{fig:cosmos-ibc} shows an example communication of a datagram from Zone 1 to Zone 3 using Cosmos' Inter-Blockchain Communications (IBC) system. A transaction on the Zone 1 blockchain generates the datagram to be communicated to Zone 3 blockchain. The datagram is included in a block on Zone 1 blockchain. A \textit{Block Commit} message containing the block header of the block containing the datagram to be communicated is signed by at least a threshold number of validators and sent to the Hub blockchain. A \textit{Packet} message containing the datagram, and a Merkle Proof proving that the datagram relates to the block, are sent to the Hub blockchain. The Hub blockchain then includes the datagram in a block on the Hub blockchain. A node on the Hub blockchain then sends a \textit{Block Commit} message to the Zone 3 blockchain with the Hub blockchain's block header for the block containing the datagram. This message is signed by at least a threshold number of validators on the Hub blockchain. Finally, the Hub blockchain sends the datagram along with a Merkle Proof in a \textit{Packet} message. The validators on Zone 3 blockchain can send a \textit{Block Commit} message and then an acknowledgement message back to the Hub blockchain. Once the Hub blockchain includes the acknowledgement in a block, validators on Zone 1 blockchain can check the acknowledgement and be sure that the transaction has been included in the Zone 3 blockchain.

The datagram sent from the Zone 1 blockchain can contain a timeout defined in terms of a block number on the Hub blockchain. If the Hub blockchain determines that a datagram has timed out, because validators on the Hub blockchain have not seen an acknowledgement from Zone 3 blockchain, they can generate a \textit{timeout datagram}, include it in the Hub blockchain, send a \textit{Block Commit} and then a \textit{Timeout} message to the Zone 1 blockchain.

Cosmos' crosschain consensus relies on the trust between the Zone and Hub blockchains, and the threshold number of validators that need to sign a message used for each of them. The messages contained in the \texttt{Block Commit} messages are trusted based on this. In turn, datagrams are shown to have been included in blocks by presenting Merkle Proofs in \textit{Packet} messages. 

The system does not provide guaranteed atomic behaviour. For example, there is no guarantee that a datagram is included in both the Zone 1 and Zone 3 blockchains. There is also no guarantee that the Zone 3 blockchain will be willing to submit acknowledgements to the Hub blockchain. The Hub blockchain should produce a timeout datagram, though there is no certainty that this will occur.

The liveness of the system depends on the liveness of the individual Zone and Hub blockchains. 

The Hub blockchain is provided rate control and protected against Denial of Service (DoS) attacks flooding the system with crosschain transactions by charging for each transaction committed to the Hub blockchain using Cosmos' digital currency, the \textit{Atom}. This usage of the Atom could lead to issues for users who do not have adequate Atoms to pay for blocks to be included in the Hub blockchain. In particular, acknowledgements posted by blockchains receiving crosschain transactions may not be able to be posted to the Hub blockchain if accounts on the receiving blockchain do not have adequate Atoms.

The system relies on Zone blockchain validators fully trusting Hub blockchain validators to act correctly. That is, the Hub blockchain acts as a centralisation point for the entire system of blockchains.

Cosmos~\cite{cosmos2016} documentation indicates that cross-zone messages can be end-to-end encrypted which would provide confidentiality. This might allow the cross-zone message system to be used between two Private Blockchain Zones.

\subsection{Polkadot}
\label{sec:polkadot}
Polkadot~\cite{polkadot2016, polkadot2019} proposes a multi-chain network consisting of \textit{Relay Chains}, \textit{Parachains} and \textit{Bridges}. Relay Chains provide shared consensus for all Parachains. Parachains receive and process transactions. Bridges provide a mechanism for transactions to be routed to non-Polkadot blockchain systems. Note that, despite the name, Relay Chains do not relay messages between Parachains or Bridges. 

There are two main roles that participants play in the Polkadot ecosystem: \textit{Collator} and \textit{Validator}. Collators collect transactions on Parachains, propose blocks and provide zero knowledge non-interactive proofs proving the transactions result in valid state changes to the Validators. Groups of Validators ratify Parachain blocks and publish them to the Parachain. The Validators seal the Parachain block headers to the Relay Chain. The Validators are randomly assigned to Parachains, with the assignment changing regularly. Validators use a PoS consensus algorithm to provide a shared consensus for all Parachains. Supportive roles are performed by \textit{Nominators} and \textit{Fishermen}. Nominators provide funds to Validators they trust to execute the PoS consensus. Fishermen observe the Parachains and submit fraud proofs to Validators.

Cross-Parachain transactions are identical to typical transactions from external accounts. A transaction on one Parachain results in a message being placed in an outbound queue by a Collator on that chain. A Collator for the target Parachain will gather messages destined for the Parachain and submit them to the Parachain's incoming queue. The message is then processed as a transaction on the destination Parachain. The messages are trusted by the destination Parachain as the proof that the message relates to a transaction on the originating Parachain can be submitted, and the transaction can be proven to have been included in a block.

Transactions from Polkadot to Ethereum via a Bridge are achieved by submitting transactions to a special multi-signature Ethereum contract. The signers of the multi-sig wallet are likely to be Validators. Transactions from Ethereum to Polkadot are achieved by calling into a special Ethereum contract which writes an event to the Ethereum event log. This event is interpreted as the outward-bound call. This technology was ``archived" in September 2020~\cite{parity-bridge-github}. Transactions between Bitcoin and Polkadot can be achieved using the XCLAIM protocol~\cite{xclaim, interlay-interbtc}. This instantiation of the XCLAIM protocol uses the same techniques as BTC Relay, but with the Polkadot blockchain.

Despite Polkadot appearing to be a multi-blockchain system, the system is in reality a multi-shard system with shared consensus. As such, for Cross-Parachain transactions, there is no cross blockchain consensus. That is, there is no need for one blockchain to prove to another blockchain that an event occurred or a transaction has been included in a block, as all nodes in the network are using the same blockchain. This usage of a shared consensus algorithm greatly simplifies Cross-Parachain communications.

If a Parachain was to be considered a separate chain, then the requirement for randomly allocated Validators to view information on the Parachain is incompatible with the concepts of private blockchains which need to restrict the list of participants.

\subsection{Atomic Crosschain Transactions for Ethereum Private Sidechains}
\label{ACT}
\label{sec:atomic-crosschain}
Atomic Crosschain Transactions~\cite{robinson2019b, robinson2020-cross-perf, crosschainwhitepaper} for Ethereum Private Sidechains~\cite{robinson2018a} and private Ethereum blockchains are a technology that allows for crosschain function calls that are both synchronous and atomic. This technology uses four key concepts to provide atomic crosschain function calls: committing to a call tree and subsequently verifying that actual parameter values match expected values, coordination blockchains, threshold signatures, and contract locking. The technology requires changes to the blockchain platform software to operate.

\subsubsection{Call Tree Commitment and Verification}
Atomic Crosschain Transactions are special nested Ethereum transactions in which the most nested transactions are first signed, and the successive encapsulating transactions are signed. The act of creating these nested transactions commits the Atomic Crosschain Transaction to a specific call tree. Compliance with the committed call tree is verified by comparing the actual parameter values with the expected values in the call tree. 

\begin{figure}
  \includegraphics[width=\linewidth]{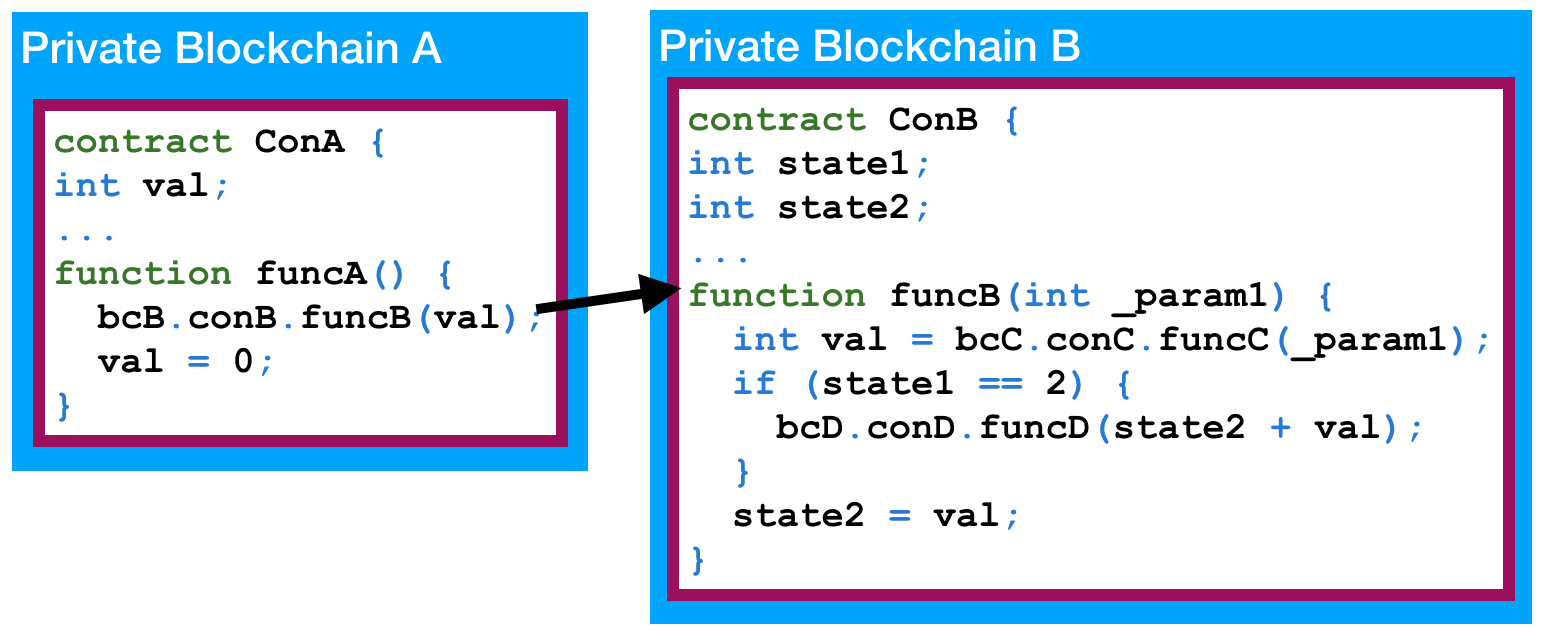}
  \caption{Function Calls across Blockchains}
  \label{fig:functioncall}
\end{figure}

When a function executes in the Ethereum Virtual Machine~\cite{wood2016a} function parameter values and stored state combine to form the actual values of variables during execution. For example, consider \texttt{funcB} in contract \texttt{ConB} on \texttt{Private Blockchain B} in Figure~\ref{fig:functioncall}. Assuming \texttt{\_param1} is \texttt{1}, \texttt{state1} is \texttt{2}, \texttt{state2} is \texttt{4}, and that \texttt{funcC} returns the value \texttt{6}, then function \texttt{funcC} will be called with the parameter value \texttt{1}, and function \texttt{funcD} will be called with the parameter value \texttt{10}. To execute this as part of a Crosschain Transaction, signed nested transactions need to be created with the appropriate parameter values. 

Execution of a transaction or view fails if the actual parameter value passed to a function does not match the value in the signed nested transaction. The parameter value in the signed nested transaction is the value the application developer expected to be passed to the function. The expected value can be determined at time of nested transaction creation using program simulation. 

\subsubsection{Blockchain Signing and Threshold Signatures}
The Atomic Crosschain Transaction system uses BLS Threshold Signatures~\cite{bls2004, bls-threshold, bls-threshold-youtube} to prove that information came from a specific blockchain. For example, nodes on one blockchain can be certain of results returned by a node on another blockchain for a function call, as the results are threshold signed by the validator nodes on the other blockchain. Similarly, validator nodes on a blockchain can be certain that validator nodes on another blockchain have mined a nested transaction, locked contracts and are holding the updated state as a provisional update because validator nodes sign a special \textit{Subordinate Transaction Ready} message indicating that the nested transaction is ready to be committed.

\subsubsection{Crosschain Coordination}
\textit{Crosschain Coordination Contracts} exist on \textit{Coordination Blockchains}. They allow validator nodes to determine whether the provisional state updates related to the Atomic Crosschain Transaction should be committed or discarded. The contract is also used to determine a common timeout for all blockchains, and as a repository of Blockchain Public Keys. 

When a user creates a Crosschain Transaction, they specify the Coordination Blockchain and Crosschain Coordination Contract to be used for the transaction, and the timeout for the transaction in terms of a block number on the Coordination Blockchain. The validator node that the user submitted the Atomic Crosschain Transaction to (the Originating Node) works with other validator nodes on the blockchain to sign a \textit{Crosschain Transaction Start} message. This message is submitted to the Crosschain Coordination Contract to indicate to all nodes on all blockchains that the Crosschain Transaction has commenced. 

When the Originating Node has received Subordinate Transaction Ready messages for all nested transactions, it works with other validator nodes to create a \textit{Crosschain Transaction Commit} message. This message is submitted to the Crosschain Coordination Contract to indicate to all nodes on all blockchains that the Crosschain Transaction has completed, and all provisional updates should be committed. If an error is detected, then a \textit{Crosschain Transaction Ignore} message is created and submitted to the Crosschain Coordination Contract to indicate to all nodes on all blockchains that the Crosschain Transaction has failed, and all provisional updates should be discarded. Similarly, if the transaction times-out, all provisional updates will be discarded.

\subsubsection{Contract Locking and Provisional State Updates}
The act of mining a nested transaction and including it in a blockchain locks the contract. The contract is unlocked when the Crosschain Coordination Contract is in the \textit{Committed} or \textit{Ignored} state, or when the block number on the Coordination Blockchain is greater than the Transaction Timeout Block Number. The Crosschain Coordination Contract will change from the \textit{Started} state to the \textit{Committed} state when a valid Crosschain Transaction Commit message is submitted to it, and it will change from the \textit{Started} state to the \textit{Ignored} state when a valid Crosschain Transaction Ignore message is submitted to it. 

Ordinarily, all nodes will receive a message indicating that they should check the Crosschain Coordination Contract when the contract can be unlocked. When a node first processes a transaction, it will set a local timer which should expire when the Transaction Timeout Block Number is exceeded. If the node has not received the message by the time the local timer expires, the node checks the Crosschain Coordination Contract to see if the Transaction Timeout Block Number has been exceeded and whether the updates should be committed or ignored.

\subsubsection{Analysis}
Atomic Crosschain Transactions require a set of validators to threshold sign information. The threshold signing mechanism does not reveal any information about the validators who signed or the threshold number of validators that was needed to sign. 

The identity of the node that submits the \textit{Start}, \textit{Commit}, and \textit{Ignore} messages is exposed when the messages are submitted to the Crosschain Coordination Contract. This exposure could be mitigated by using a new randomly generated identity on the Coordination Blockchain for each submission to the contract. The Crosschain Coordination Contract trusts the messages based on the threshold signatures, and not on the transaction signature.

Due to the simple fail-if-locked locking scheme, the system can not have \textit{dead locks}, though could suffer from \textit{live lock}.

Due to the global per-transaction timeout offered by the Coordination Blockchain, the system will not have liveness issues. Once a Crosschain Transaction is committed, ignored, or times-out, all contracts are unlocked. If nodes on a blockchain refused to process a \textit{Signalling Transaction}, then contracts that were part of the crosschain transaction on the blockchain would remain locked. However, this situation is analogous to nodes refusing to produce any more blocks, and hence not related to crosschain consensus.

The system does not allow for partial updates. Errors are returned if a user attempts to read values from a locked contract. As such, this methodology is a \textit{safe} crosschain consensus mechanism. 

The main drawback of Atomic Crosschain Transactions is that it requires changes to the Ethereum Client software. As such, this technology is not suitable for applications where the Ethereum Client software can not be modified.

\subsection{General Purpose Atomic Crosschain Transactions}
\label{sec:gpact}
The General Purpose Atomic Crosschain Transaction (GPACT) protocol~\cite{robinson2020general,gpact-ieee-poster} is a technology that allows function calls across blockchains that either update or discard state updates on all blockchains. This protocol leverages the ideas developed for Atomic Crosschain Transactions (see Section~\ref{ACT}). Applications determine parameter values by simulating a call tree, then commit to the call tree. Segments of the call tree execute successfully if the call tree can be executed using the parameter values determined in the simulation. Values within a contract's state are locked if a call segment updates those values. These provisional updates are applied to all locked contracts if the overall crosschain transaction is successful and discarded if not.

The protocol relies on communicating messages between blockchains such that they are trusted. In the reference implementation of the protocol~\cite{gpact-github} the crosschain messaging has been achieved by having a set of \textit{Attestors} for each blockchain sign Ethereum Events~\cite{wood2016a}. The Attestors cooperate to sign the events. Users request the signed events they are interested in from one of the Attestors. The events are trusted on destination blockchains if at least a threshold number of Attestors have signed the event. 

The first step to execute the protocol is to determine the expected parameter values for entry point function calls for contracts on blockchains. The application using the protocol needs to fetch state from the contracts and then execute a simulation of the contract code. To commit to the call tree, a transaction is submitted to the Root Blockchain's \textit{Crosschain Control Contract}'s \textit{Start} function. The call tree is then emitted as part of a \textit{Start Event}.

Once the Start Event has been emitted, \textit{Segment} functions on the Crosschain Control Contract on blockchains that make up the call tree of the overall transaction are  called to request a function on a contract be called as part of a crosschain function call. The signed Start Event and an indicator of where the function call lies in the call tree are submitted as parameters to prove that this function is part of the crosschain function call. Additionally, a set of signed \textit{Segment Events} containing function call return results are passed in to prove that subordinate function calls on other blockchains have been called and have returned certain results. 

When the application contract code executes, if a crosschain function call is encountered, the application code calls the \textit{CrossCall} function in the Crosschain Control Contract, passing in the actual parameters of the function call. The  CrossCall function compares the actual parameter values with the expected parameter values as specified in the Start Event. A Segment Event is emitted to publish the return result or error result of the function, and the list of contracts containing locked values. 

The \textit{Root} function is called on the Root Blockchain to call the entry point function call for the call tree. The Root function call has similar parameters to the Segment function call, taking a signed Start Event and a set of signed Segment Events. Similar to the Segment function call, expected and actual parameter checking is completed along with contract locking.

If a Root function call completes successfully, any locked contracts on the Root Blockchain are unlocked and provisional state updates are committed. A \textit{Root Event} is emitted indicating that all provisional updates on blockchains should be committed. The Root function emits a Root Event indicating that all updates on all other blockchains should be discarded if any of the Segment functions returned error results or if an error occurs while executing the entry point function call. 

If the timestamp of the most recent block on the Root Blockchain is after the timeout in the Start Event, then any account can submit a transaction that calls the Root function to cancel the crosschain function call. In this situation a Root Event is emitted that indicates that all updates on all other blockchains should be discarded.

The \textit{Signalling} function is called on blockchains that have updates that need to be committed or discarded. The signed Root Event and signed Segment Events for the blockchain are passed in as parameters. The Root Event indicates whether updates should be committed or discarded. The Segment Events contain the list of contracts containing values that need to be unlocked.

The GPACT protocol guarantees safety by committing to a call tree, checking for correct crosschain execution, locking values, and committing or discarding those values depending on the overall execution of the crosschain transaction. That is, contracts on separate blockchains will have a consistent state by all applying or all discarding changes. Liveness is assured by the timeout feature of the protocol, as all contracts will eventually be unlocked.

\section{Pinning Techniques}
\label{sec:pinning}

\subsection{Merge Mining}
\label{sec:merged}
Merge Mining~\cite{namecoin2015,  merged-mining-2011, merged-mining2014}
is a technique in which the block hash of a low hashing power public blockchain, such as NameCoin, is included in a more secure higher hashing power blockchain, such as Bitcoin. In this scenario, the Bitcoin miners must validate NameCoin transactions prior to including the Block Hash in a transaction on the Bitcoin network. The mined transaction can then be included in both the Bitcoin and NameCoin blockchains. 

Merged mining relies on both blockchains using the same consensus algorithm and assumes that all transactions can be viewed by both blockchains. As such, this technique is not usable in a private blockchain scenario where transactions must not be revealed outside the blockchain.

The liveness of a merge mined blockchain relies on the miners being sufficiently incentivised to include blocks from the merge mined blockchain in the higher hashing power blockchain. The safety of the technique relies on a variety of miners on higher hashing power blockchain choosing to include blocks. For example, as of 2015, NameCoin has only been mined by one miner~\cite{namecoin2015}.

\subsection{Tethered Blockchains: Block Hash Posting}
\label{sec:tethered}
The Kaledio team developed Tethered Permissioned Private Chains~\cite{kaleido-relay, kaleido} to reduce the risk of state reversion occurring by posting the state of the blockchain onto Ethereum MainNet. In Kaleido's system, a trusted entity may be used to submit a block hash on behalf of blockchain participants, thus keeping the participants secret. 

The rate of transactions on the consortium chain is revealed on Ethereum MainNet due to the number of posted pins. Additionally, the solution lacks a method of contesting a pin posted to Ethereum MainNet.

\subsection{Anonymous Block Hash Posting}
\label{sec:anon-pin}
\begin{figure}
  \includegraphics[width=\linewidth]{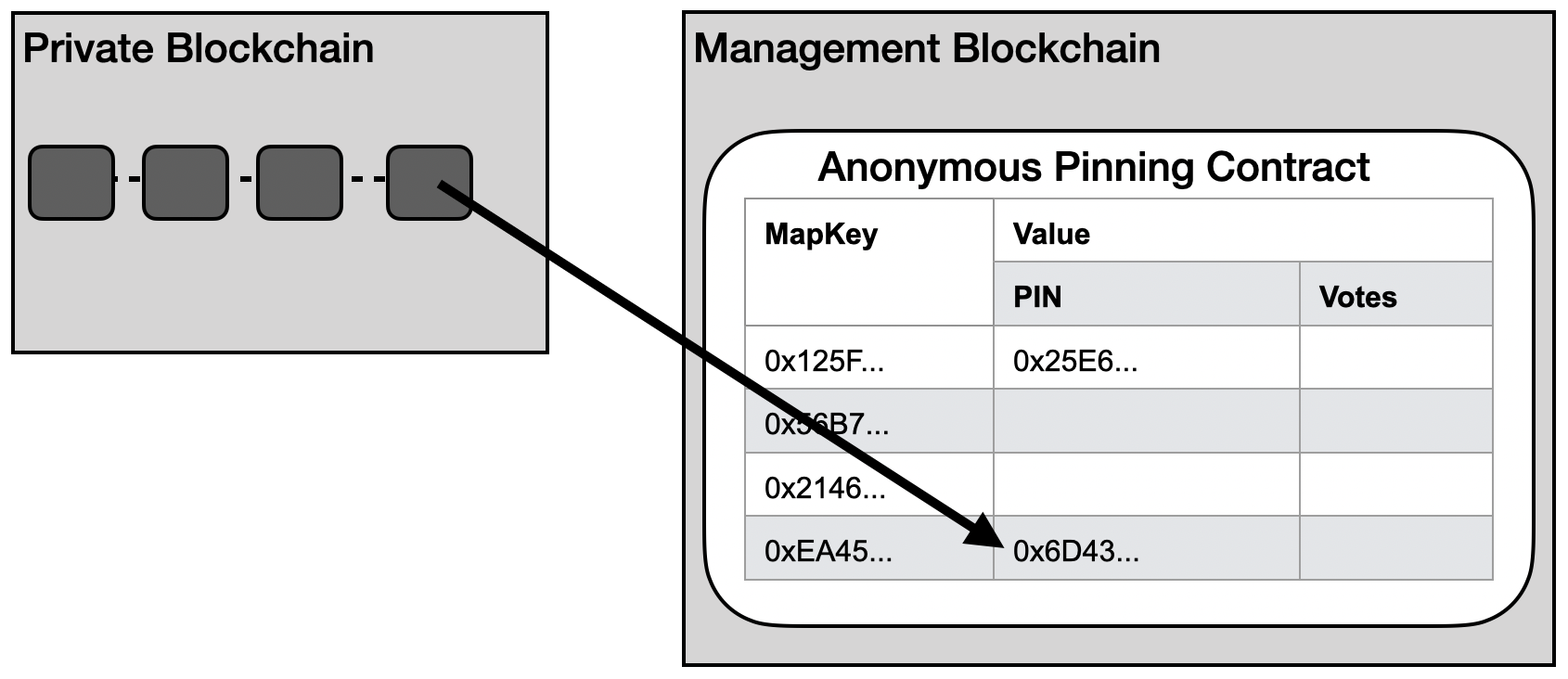}
  \caption{Anonymous Pinning Map}
  \label{fig:anon-pinning}
\end{figure}

Robinson and Brainard proposed Anonymous State Pinning~\cite{anonpinning} as a method of posting block hashes of a private blockchain to a \textit{Management} blockchain without revealing the identities of participants of the private blockchain or the rate of transactions, while allowing posted block hashes to be contested. Pins (block hashes) are posted to certain entries in a map as shown in Figure~\ref{fig:anon-pinning}. 

The value of the \texttt{MapKey} value is calculated using Equation~\ref{equ-mapkey}, where \texttt{KECCAK} is a message digest algorithm, Private Blockchain Identifier (\texttt{PBI}) identifies the blockchain, \texttt{Pin\textsubscript{t-1}} is the previous pin value, \texttt{PRF} is a pseudo random function which is seeded with a Private Blockchain Secret.

\begin{equation}
\label{equ-mapkey}
MapKey\textsubscript{t} = KECCAK\textnormal{-}256( PBI, Pin\textsubscript{t-1}, PRF(t))
\end{equation}
Participants of a private blockchain observe the pinning map at the MapKey\textsubscript{t} address corresponding to the next pin, waiting for the next pin to be posted to that entry in the map. When the pin value is posted, they check that the posted pin matches their understanding of the most recent block hash of the private blockchain. If the values do not match, then participants should contest the pin. To contest the pin, Pin\textsubscript{t}, which is at MapKey\textsubscript{t}, they submit to the contract: MapKey\textsubscript{t-1}, PRF(t), and the PBI.

Submitting the previous value of the MapKey, MapKey\textsubscript{t-1}, allows the contract to fetch from its own storage the value of the previous pin, Pin\textsubscript{t-1}. The contract can then calculate the MapKey of the contested pin, MapKey\textsubscript{t}, by combining Pin\textsubscript{t-1}, PRF(t) and the PBI using Equation \ref{equ-mapkey}. Given the submitter of the transaction knows the PRF(t) which combined with the PBI links the previous MapKey, MapKey\textsubscript{t-1}, and the calculated MapKey, MapKey\textsubscript{t}, it implies that both MapKeys correspond to pins for the private blockchain denoted by PBI. The further implication of knowing PRF(t) is that the transaction submitter has access to the Private Blockchain Secret, which implies that they are a member of the private blockchain.

To hide the identity of participants, \textit{masked participants} are represented by a salted hash or their account number. \textit{Masked participants} may unmask themselves at any time to become \textit{unmasked participants} by presenting their secret salt value. Only \textit{unmasked participants} can vote on whether a contested pin is valid.

As multiple blockchains can use the same map, it is difficult for observers to determine the rate of pinning for a particular blockchain, assuming one account is used to submit transactions for multiple blockchains. 

The liveness of the system depends on pins being posted when needed. The safety of the system depends on all participants being able to observe the management chain at all times so that they can contest pins when needed. This requirement of being online at all times is onerous. It may not be able to be fulfilled in all deployments.

\subsection{Threshold Signed Block Hash Posting}
\label{sec:threshold-signed}
The validators of a blockchain could collaborate to use BLS Threshold Signatures~\cite{bls2004, bls-threshold, bls-threshold-youtube} to threshold sign block hashes that are submitted to a \textit{management} blockchain. In this scheme, the public key related to the private key shares would be stored on the  \textit{management} blockchain. The signed pins are verified when they are submitted to the contract used to store the pins, thus ensuring only valid pins signed by a majority of validators are accepted. An alternative to using BLS Threshold Signing would be to have multiple parties submit signed block hashes. If a threshold number of signers submitted the block hash, it would be deemed valid.

The advantage of this technique over the \textit{Anonymous Block Hash Posting} is that validators do not need to monitor the management blockchain as only valid pins are accepted. A disadvantage of this technique is that it is more computationally expensive both on the private blockchain side, having to threshold sign the pin, and in the smart contract on the management blockchain, where a BLS signature needs to be verified.

\section{Discussion}
\label{sec:discussion}
\label{sec:comparison}
Tables~\ref{table:comparison1}, ~\ref{table:comparison2}, and ~\ref{table:comparison3} lists the crosschain communications protocols described in this paper, summarising the trust assumptions for each protocol, whether the protocol provides atomic updates across blockchains, and whether the implementation of the protocol requires changes to the blockchain platform to operate. To be deemed to deliver atomic updates a protocol must be free of safety and liveness issues.  

\begin{table*}[t]
  \centering
    \caption{Value Swap Crosschain Consensus Techniques}
  \label{table:comparison1}

    \begin{tabular}{| l || l | c | c |}
    \hline
    Crosschain                          & Trust Assumption       & Atomic    & Block-   \\
    Technology                          &                                   & Updates  & chain                     \\
                                                &                                   &                 & Changes \\
                                                &                                   &                 & Required \\
       \hline
HTLCs                                       & Trustless                   & $\times$  & No    \\
(Sec.~\ref{sec:techniques:hash-timelock-contracts}) & & & \\
       \hline
Interledger                                        & Trustless                  & $\checkmark$  & Yes \& No  \\
STREAM                                   & & & \\
(Sec.~\ref{sec:interledger})       & &  & \\
       \hline
BTC Relay                                & Bitcoin's PoW and one honest Relayer           & $\times$ & No \\
(Sec.~\ref{sec:btcrelay})           & &  & \\
       \hline
XCLAIM                                    & Bitcoin's PoW and one honest Relayer           & $\checkmark$ & No \\
(Sec.~\ref{sec:xclaim})              & &  & \\
       \hline
Pegged                                     & Bitcoin and sidechain's PoW and one honest  & $\times$  & Yes \\
Sidechains                                & Relayer.                                                              &                & \\
(Sec.~\ref{sec:pegged})            & &  & \\
       \hline
Plasma                                     & Chain observers submit fraud proofs within   & $\times$  & No\\
(Sec.~\ref{sec:plasma})            & fraud window.  &  & \\
\hline
Wanchain                                 & Threshold number of Relayers honest. Slashing  & $\times$ & No\\
(Sec.~\ref{sec:wanchain}).       & is enough to keep Relayers honest.  & & \\
       \hline
  \end{tabular}
\end{table*}

\begin{table*}[t]
  \centering
    \caption{Message Passing Crosschain Consensus Techniques}
  \label{table:comparison2}

    \begin{tabular}{| l || l | c | c |}
    \hline
    Crosschain                          & Trust Assumption       & Atomic       & Block-\\
    Technology                          &                                   & Updates     & chain                 \\
                                                &                                   &                    & Changes\\
                                                &                                   &                    & Required\\
       \hline
Ion                                            & Threshold number of Relayers honest  & $\times$  & No \\
(Sec.~\ref{sec:ion}) & &  & \\
       \hline
ChainBridge                             & Threshold number of Relayers honest  & $\times$ & No \\
(Sec.~\ref{sec:chainbridge}) & & & \\
       \hline
Celo Optics                              & One observer slashes dishonest Relayers and & $\times$ & No  \\
(Sec.~\ref{sec:celo})                & application observers blocks invalid messages. &  & \\
       \hline
Cosmos                                    & Threshold number of signers on each chain & $\times$ & Yes \& No \\
(Sec.~\ref{sec:cosmos})           &  honest. & & \\
       \hline
Polkadot                                   & Relay chain operated honestly. & $\times$  & Yes \\
(Sec.~\ref{sec:polkadot})          & Note: Cross-shard technique. & & \\
       \hline
Atomic                                     & Threshold signing of messages. & $\checkmark$ & Yes \\
Crosschain                              &  & & \\
(Sec.~\ref{sec:atomic-crosschain}) & & & \\
       \hline
GPACT                                    & Depends on underlying message transfer & $\checkmark$ & No \\
(Sec.~\ref{sec:gpact})             &  technique. & & \\
       \hline
  \end{tabular}
\end{table*}

\begin{table*}[t]
  \centering
    \caption{Pinning Crosschain Consensus Techniques}
  \label{table:comparison3}

    \begin{tabular}{| l || l | c |}
    \hline
    Crosschain                          & Trust Assumption        & Block- \\
    Technology                          &                                                 & chain          \\
                                                &                                                 & Changes \\
                                                &                                                 & Required \\
       \hline
Merge                                     & Consensus of chain merged to.      &  Yes\\
Mining                                    &    & \\
(Sec.~\ref{sec:merged})   & & \\
       \hline
Tethered                                  & All validators are honest.   &  No \\
Blockchains                             &   &  \\
(Sec.~\ref{sec:tethered})   & & \\
       \hline
Anonymous                            & Observers contest invalid block hashes &  No\\
Pinning                                    &   &  \\
(Sec.~\ref{sec:anon-pin})   & & \\
       \hline
Threshold                               &  Threshold signed block hashes.    &  No \\
Pinning                                    &    &    \\
(Sec.~\ref{sec:threshold-signed})   & & \\
       \hline
  \end{tabular}
\end{table*}

HTLCs and Interledger's STREAM protocol rely on hash commitments and preimages. Interledger overcomes the potential griefing issues of HTLCs by sending payments incrementally. However, users can still withdraw from the protocol prior to completing the transfer. Both protocols can use collateral deposits to reduce the risk of griefing and exit prior to completing transfers. These trustless protocols are appropriate for permissionless blockchain platforms. 

BTC Relay, XCLAIM and Pegged Sidechains each assume the hashing power of the source blockchain is such that attackers can not mount a 51\% attack~\cite{fifty-one-percent-attack}, and the PoW algorithm can be confirmed programmatically on the destination blockchain. Blockchains other than Bitcoin and Ethereum MainNet do not have high enough hashing power to prevent 51\% attacks~\cite{robinson2020-mainnet}. Ethereum MainNet is likely to switch to PoS in early 2022. As such, the protocols are limited to use with Bitcoin.

Wanchain, Ion, ChainBridge, Cosmos, Atomic Crosschain Transactions for Ethereum Private Sidechains, and GPACT rely on threshold signing to achieve crosschain consensus. For a permissioned blockchain, the validators of the blockchain could be used as the signers. In this case, the crosschain consensus is as trustworthy as the permissioned blockchain itself. Similarly, if the transfer was from a permissionless blockchain to permissioned blockchain, the validators of the permissioned blockchain could be used. 

Wanchain and Celo Optics combine signing with staking and slashing. This crypto-economic incentivisation makes them consistent with permissionless blockchains as permissionless blockchains rely on crypto-economic incentivisation for their security. A challenge for each protocol is ensuring any potential gain due to bad behaviour is less than the amount that can be slashed.

A challenge with crosschain protocols is to ensure operators are incentivised adequately, even if there are only a small number of transactions. For example, BTC Relay stopped operation in February 2018~\cite{btc-relay-etherscan} because not enough users were confirming transactions and Relayers were only compensated when users confirmed transactions. Wanchain overcomes this issue by paying Storemen node operators to operate their nodes. 

Plasma and Anonymous Pinning allow information transferred across chains to be challenged. Plasma allows users to present proofs, proving information is invalid. Anonymous Pinning allows data to be optimistically pinned, and challenged by a threshold number of users saying information is invalid. 

Atomic Crosschain Transactions for Ethereum Private Sidechains and GPACT provide crosschain function call capabilities. They check that processing on all blockchains is successful before finalising updates on all blockchains. This means that state updates across blockchains are consistent. Protocols that only provide crosschain messaging can not guarantee this crosschain consistency.

Polkadot though appearing to be a crosschain protocol is in fact a cross-shard protocol. Having a trusted Relay Chain that the Parachains integrate with means that no crosschain consensus needs to be created. Not requiring crosschain consensus simplifies the system. 

Most crosschain protocols act as blockchain applications. Protocols that require changes to blockchain platform software have limited applicability as blockchain client software developers do not want to integrate complex crosschain software into their products.

\section{Future Directions}
\label{sec:future}
Crosschain consensus is a new field with many areas to be researched further. Trustless crosschain messaging for permissionless blockchains needs to be researched further. Staking and slashing approaches have been proposed. However, how the amount staked relates to the value being transacted using the crosschain transactions needs more thought. 

In PoS blockchains, validators stake value and sign block headers. The possibility of leveraging the stake deposited by validators for securing the blockchain to also secure the crosschain communications should be investigated. 

For threshold signing in the context of PoS, an unsolved challenge is to ensure that signers in the verification contract on the destination blockchain match the PoS voting committee membership. In PoS schemes, the membership of the voting committee changes as part of the PoS protocol. The set of valid signers in the verification contract on the destination blockchain needs to change in sync with the changing voting committee membership.

Collateral deposits have been proposed for a variety of crosschain protocols. An area that needs further research is the impact of high and changing transaction fees on these protocols. That is, if a transaction fee is large relative to a collateral deposit, it may not make economic sense to use collateral deposits.

Crosschain function call approaches need further research to determine whether more efficient approaches can be found. Open research questions include, can \textit{gas} efficient approaches be determined and can locking strategies be improved?

New types of blockchain technologies called Optimistic and Zero Knowledge Rollups~\cite{rollups-incomplete-guide} are being developed. These new technologies have different properties to traditional blockchains, and hence are likely to be best suited to new crosschain consensus approaches. In particular, the ability to prove that a Relayer has misbehaved may be challenging. Relayers cannot be slashed without the ability to prove on-chain or on-rollup that the relayer has misbehaved.

\section*{Acknowledgments}
This research has been undertaken whilst I have been employed full-time at ConsenSys Software. I acknowledge the support of my PhD co-supervisors Dr Marius Portmann and Dr David Hyland-Wood. I thank Dr Sandra Johnson with whom I co-authored an early paper on crosschain interoperability that was part of the inspiration for this paper. I thank Dr Catherine Jones for her continued support and astute review of this paper. I thank the anonymous reviewers for their comprehensive review of this paper.

\ifARXIV

\bibliographystyle{IEEEtran}
\bibliography{IEEEabrv,ref}

\end{document}

\else

\bibliographystyle{elsarticle-num}
\bibliography{ref}

\end{document}
\endinput

\fi